\documentclass[useAMS,usenatbib,usegraphicx]{mn2e}
\usepackage{bm}
\usepackage{times} 

\begin{document}
\title[Dust shattering and coagulation in interstellar turbulence]
{Shattering and coagulation of dust grains in interstellar turbulence}
\author[H. Hirashita \& H. Yan]{Hiroyuki Hirashita$^{1}$\thanks{E-mail:
    hirashita@asiaa.sinica.edu.tw} and Huirong Yan$^{2,3}$
\\
$^1$ Institute of Astronomy and Astrophysics,
 Academia Sinica, P.O. Box 23-141, Taipei 10617, Taiwan \\
$^2$ Canadian Institute for Theoretical Astrophysics, 60 St.\
George Street, Toronto, ON M5S 3H8, Canada\\
$^3$ University of Arizona, Steward Observatory, 933 N
Cherry Avenue, Tucson, AZ 85721, USA
}
\date{2008 December 15}
\pubyear{2008} \volume{000} \pagerange{1}
\twocolumn

\maketitle \label{firstpage}
\begin{abstract}
We investigate shattering and coagulation of dust grains in
turbulent interstellar medium (ISM). The typical velocity of
dust grain as a function of  grain size has been calculated
for various ISM phases based on a theory of grain dynamics
in compressible magnetohydrodynamic turbulence. In this
paper, we develop a scheme of grain shattering and coagulation
and apply it to turbulent ISM by using the grain velocities
predicted by the above turbulence theory. Since large grains
tend to acquire large velocity dispersions as
shown by earlier studies,
large grains tend to be shattered.
Large shattering effects are indeed seen in warm ionized medium
(WIM) within a few Myr for grains with radius $a\ga 10^{-6}$ cm.
We also show that shattering in warm neutral medium (WNM) can
limit the largest grain size in ISM
($a\sim 2\times 10^{-5}~\mathrm{cm}$).
On the other hand, coagulation tends to modify small
grains since it only occurs when the grain velocity is
small enough.
Coagulation significantly modifies the grain size
distribution in dense clouds (DC), where a large fraction
of the grains with $a<10^{-6}$ cm coagulate in 10 Myr.
In fact, the
correlation among $R_V$, the carbon bump strength, and the
ultraviolet slope in the observed Milky Way extinction
curves can be explained by the coagulation in DC.
It is possible that the grain size distribution in the
Milky Way is determined by a combination of all the
above effects of shattering and coagulation. Considering
that shattering and coagulation in turbulence are effective
if dust-to-gas ratio is typically more than
$\sim 1/10$ of the Galactic value, the regulation mechanism
of grain size distribution should be different
between metal-poor and metal-rich environments.
\end{abstract}

\section{Introduction}

Dust grains absorb stellar ultraviolet (UV)--optical light
and reprocess it into far-infrared (FIR), thereby affecting
the energetics of interstellar medium (ISM)
\citep[e.g.][]{hirashita02}. For this process concerning
the interaction between grains and radiation, the optical
properties of grains are important. The grain optical
properties are determined not only by dust species
but also by the grain size \citep[e.g.][]{draine84}.
In fact, the grain species and size distribution are
derived from the observed interstellar extinction curve
\citep[][hereafter MRN]{mathis77}. These grain properties
also affect the FIR spectrum of dust emission
\citep[e.g.][]{takeuchi05}.

The grain size distribution is known to be affected by
various processes in the interstellar space. Grains are
supplied from stars at their death \citep{gehrz89}
with a certain grain size distribution
\citep{dominik89,todini01,nozawa03,bianchi07,nozawa07}.
These grains dispersed from stars are processed in the
interstellar space. They grow in dense environments such
as molecular clouds by grain-grain coagulation
 {\citep*{chokshi93}} and accretion of heavy
elements \citep{spitzer78}. They are also destroyed by
supernova shocks by gas-grain sputtering
and by grain-grain collision (shattering)
\citep{dwek80,tielens94,borkowski95}.
In particular,  {\citet[hereafter JTH96]{jones96}}
show that the grain size
distribution can be significantly modified by shattering.
Such a change of grain size distribution would
significantly modify the extinction curve and the
infrared spectral energy distribution of dust emission.

Potential importance of shattering and coagulation
in ISM has often been pointed out, since relative
velocity between grains is naturally expected if ISM is
turbulent
\citep{kusaka70,volk80,draine85,ossenkopf93,lazarian02}.
Because turbulence is ubiquitous in ISM
 {\citep[e.g.][]{mckee07}}, the relative grain
motion induced by turbulence is of
general importance in the grain evolution in ISM.
Moreover, the ISM is known to be magnetized
\citep{arons75}. Thus, dust motion in magnetohydrodynamic
(MHD) turbulence should be considered.
\citet[hereafter YLD04]{yan04} calculate the relative grain
velocity in compressible MHD turbulence, taking into
account gas drag (hydrodrag) and gyroresonance. The basis
of their theory
can be seen in \citet{lazarian02} and \citet{yan03}.
According to their results, grains can be accelerated
to a velocity larger than a few km s$^{-1}$ in diffuse
medium by gyroresonance. At such a high velocity,
grains can be shattered (JTH96).
On the other hand, small grains can obtain
velocities small enough for coagulation to occur,
especially in dense medium (YLD04).

The size distribution of grains processed in various ISM
phases was investigated by \citet{odonnell97}. Their models
incorporate coagulation by turbulent motion in clouds,
accretion of gas-phase metals onto grains, and shattering
and sputtering in interstellar shocks. They also take into
account the
phase exchange of the ISM. They show that both the observed
extinction curve and the observed depletion of refractory
elements are reproduced by considering phase exchange
among diffuse clouds, warm neutral intercloud gas,
and molecular clouds. In their models, shattering
mainly occurs in ISM shocks. However, \citet{yan03} shows
that grains can be accelerated to velocities large
enough for shattering in MHD turbulence. Thus, it is
worth focusing on the effect of turbulence on the grain
size distribution. In addition, the treatment of shattering
can be revised by including the framework of JTH96 to take
into account the velocity dependence of fragment production.
It is also
useful to compare the difference between the size
distribution modified by supernova shocks as treated
by JTH96 and that by interstellar turbulence as examined
in this paper.

The aim of this paper is to examine quantitatively whether
or not shattering and coagulation in turbulent ISM really
modify the grain size distribution. We focus on the effects
of MHD turbulence as treated in \citet{yan03}.
The other types of dust processing such
as shattering and sputtering in supernova shocks and dust
condensation in stellar mass loss are not
treated in this paper, in order to make our discussions
concentrated and clear. For an observational comparison,
we adopt the Milky Way extinction curve following
\citet{odonnell97}.

This paper is organized as follows. First, in
Section~\ref{sec:model}, we describe the model adopted
to treat shattering and coagulation in turbulent ISM.
Then, in Section~\ref{sec:result}, we overview the
results.
In Section~\ref{sec:discussion}, we discuss our results,
focusing on the regulation mechanism of the grain size
distribution in turbulent ISM. Section~\ref{sec:sum}
is devoted to the summary.

\section{Model}\label{sec:model}

We consider the evolution of grain size distribution by
shattering and coagulation induced by relative grain
motions in turbulence. In our models, shattering and
coagulation are treated simultaneously. The basic
ingredients for shattering and coagulation are taken from
JTH96 and \citet{chokshi93}, respectively. We do not consider
vaporization in grain collision, since this process does
not change the grain size significantly at velocities
(at most a few tens km s$^{-1}$) achieved in turbulence.
 {Our results are most sensitive to the collision
rate between grains. Thus, first of all, the grain velocities
adopted are discussed. Then, the frameworks of
shattering and coagulation are explained.}

\subsection{Grain motion}\label{subsec:vel}

We assume spherical grains (Section \ref{subsec:timeev}).
The velocity of a grain with radius $a$ in the presence of
interstellar MHD turbulence is taken from YLD04, who
calculated the grain velocities achieved in various phases
of ISM (CNM, WNM, WIM, MC, and DC, which stand for cold
neutral medium, warm neutral medium, warm ionized medium,
molecular cloud, dense cloud, respectively). The physical
parameters which they adopted for each phase are listed
in Table \ref{tab:yan}.
For DC, they adopt two
different cases for the ionization fraction (DC1 and DC2)
to examine the effect of
the uncertainty in the cosmic-ray ionization rate.
 {We also adopt the same electrical charge of grains
as calculated in YLD04, who considered
photoelectric emission and collisions
with ions and electrons based on the hydrogen number
densities, the electron densities, and the UV radiation
fields in
Table \ref{tab:yan}. Since the values adopted for these
quantities are those usually assumed for
Galactic conditions, we expect that the calculations below
are at least reasonable for the Galactic ISM.}
Grains are accelerated through turbulence hydrodrag and
gyroresonance.  {Below we
briefly overview the models of turbulence
and gyroresonance adopted by YLD04.}

\begin{table*}
\centering
\begin{minipage}{130mm}
\caption{The parameters of idealized ISM phases in YLD04. Among
them, $T$ is the gas temperature, $n_\mathrm{H}$ is the number
density of hydrogen atoms, $n_\mathrm{e}$ is the number density
of electrons, $G_\mathrm{UV}$
is the UV intensity relative to the average local interstellar
radiation field, 
$L$ is the injection scale at which equipartition between
magnetic and kinetic energies occurs, $V$ is the effective
injection velocity at the scale $L$ (i.e.\ $V=V_\mathrm{A}$, where
$V_\mathrm{A}$ is the Alfv\'{e}n velocity), and $k_\mathrm{c}$ is the
wavenumber of the damping scale of turbulence.
CNM: cold neutral medium; WNM: warm neutral
medium; WIM: warm ionized medium; MC: molecular cloud; DC: dark cloud. }
\begin{tabular}{|c|c|c|c|c|c|c|}
\hline 
 ISM Phase &
 CNM&
 WNM&
 WIM&
 MC&
 DC1&
 DC2\tabularnewline
\hline
$T$ (K)&
 100&
 6000&
 8000&
 25&
 \multicolumn{2}{|c|}{10}\tabularnewline
$n_\mathrm{H}$ (cm$^{-3}$)&
 30&
 0.3&
 0.1&
 300&
 \multicolumn{2}{|c|}{$10^4$}\tabularnewline
$n_\mathrm{e}$ (cm$^{-3}$)&
0.03&
0.03&
0.0991&
0.03&
0.01&
0.001\tabularnewline
$G_\mathrm{UV}$&
1&
1&
1&
0.1&
0.01&
0.001\tabularnewline
$B$ ($\mu$G)&
6&
5.8&
3.35&
11&
\multicolumn{2}{|c|}{80}\tabularnewline
$L$ (pc)&
0.64&
100&
100&
1&
\multicolumn{2}{|c|}{1}\tabularnewline
$V=V_\mathrm{A}$ (km s$^{-1}$)&
2&
20&
20&
1.2&
\multicolumn{2}{|c|}{1.5}\tabularnewline
$k_\mathrm{c}$ (cm$^{-1}$)&
$7\times 10^{-15}$&
$4\times 10^{-17}$&
--- &
$4.5\times 10^{-14}$&
$5.3\times 10^{-15}$&
$5.3\times 10^{-17}$\tabularnewline
\hline
\end{tabular}\centering
\label{tab:yan}
\end{minipage}
\end{table*}

\subsubsection{Turbulence}

The velocity achieved by hydrodrag is determined by the
largest scale on which grains are decoupled from the
hydrodynamical motion, since the turbulent velocity is
larger on larger scales.
Thus, larger grains, which tend to be coupled with
larger-scale motions, are more accelerated.
 {In general, neutral medium tends to accelerate
grains less than ionized medium because of ion-neutral
collision damping of MHD turbulence. For example, the
grain velocity in DC2 is smaller than DC1 because
of the difference in ionization degree. The same reason
is applied to the larger
grain velocities achieved in WIM than in WNM.}

 {YLD04 obtain the scaling relation of turbulence
velocity based on a MHD turbulence theory developed by
\citet{cho02}, who decompose MHD fluctuations into
Alfv\'{e}n, slow, and fast
modes. Unlike hydrodynamic turbulence, Alfv\'{e}nic
turbulence is anisotropic, with eddies elongated
along the magnetic field \citep{goldreich95}. This is
because it is easier to mix the magnetic fields lines
perpendicular to the direction of the magnetic field
rather than to bend them. The energies of eddies drops with
the decrease of eddy size, and it becomes more difficult
for smaller eddies to bend the magnetic field lines.
Therefore, the eddies get more and more anisotropic
as the sizes decreases. Eddies mix the magnetic field
lines at the rate of $k_\perp v_k$, where $k_\perp$ is a
wavenumber measured in the direction perpendicular to the
local magnetic field and $v_k$ is the mixing velocity.
The energy spectrum for the perpendicular motions becomes
Kolmogorov-like, i.e.\ $v_k\propto k^{-1/3}$. On the other hand,
the magnetic perturbations propagate along the magnetic
field lines at the rate $k_\parallel V_\mathrm{A}$, where
$k_\parallel$ is the wavenumber parallel to the local
magnetic field and $V_\mathrm{A}$ is the Alfv\'{e}n
velocity. The mixing motions couple to the wavelike
motions parallel to magnetic field, giving
a critical balance condition,
$k_\perp v_k\sim k_\parallel V_\mathrm{A}$.
Thus, we obtain $k_\parallel\propto k_\perp^{2/3}$.
The fast modes follow acoustic cascade, and show isotropic energy
spectra with $v_k\propto k^{-1/4}$ 
\citep{cho02}.}

 {YLD04 assume that equal amounts of energy are transferred
into fast and Alfv\'{e}n modes when driving is on large scales. The
cascades proceed to small scales without much cross talk between
those two kinds of modes, according to the results in
\citet{cho02,cho03}. From the scaling
relations of $v_k$ with $k$, we observe that the
decoupling from fast modes usually brings larger
velocity dispersions to grains than Alfv\'{e}n
modes.}

 {In this paper, we do not consider imbalanced turbulence,
which develops under unequal energy flux from opposite directions
and has non-zero cross-helicity. Recent study by
\citet{beresnyak08} shows that the stronger wave of the Alfv\'en
modes has
smaller anisotropy, which indicates that the interaction of the grains
with the imbalanced
Alfv\'enic turbulence could be more efficient. However,
the results on the imbalanced turbulence
is far from quantitative. And there is no conclusive theory yet for the imbalanced
fast modes, which are more important for the acceleration according to
\citet{yan03}. The only study so
far \citep{suzuki07} indicates that fast modes are not so different as in the
balanced turbulence. In addition, imbalanced turbulence is applicable to
places near an energy source, e.g.\ the
vicinity of  a star, which we do not consider in this paper.  
}

 {
The conditions for ISM phases that YLD04 adopt imply that the turbulence is
super-Alfv\'{e}nic ($\delta V\ga V_\mathrm{A}$, where $\delta V$ is the
turbulence velocity). Indeed given the uncertainty of the strength of the
magnetic field in the ISM, we do not know whether the conjecture is universal.
However,  our results for shattering and coagulation will not be concerned
sensitively to the above debate on sub/super-Alfv\'{e}nic turbulence. The
reason is what follows. If the turbulence is sub-Alfv\'enic, the turbulence is
weak. The weak turbulence has only limited inertial range. Moreover, it is the
fast modes that dominate the acceleration of dust as demonstrated by
\citet{yan03}. And there
has been study showing that fast modes in weak turbulence are similar to fast
modes in strong turbulence apart from the modes in the narrow cone around the
$\bmath{k}$ vector in Fourier space \citep{chandran05}.
When the cascades proceed down to the scale where the critical balance
$k_\bot v_k\sim k_\|V_\mathrm{A}$ is reached, turbulence becomes strong
\citep{lazarian99}. The coherence length of the strong turbulence
$LM_\mathrm{A}^2$, is also the correlation length of the turbulence magnetic
field, where
$L$ is the injection scale of the turbulence and
$M_\mathrm{A}\equiv \delta V/V_\mathrm{A}$ is the
Alfv\'enic Mach number. It is unlikely that $M_\mathrm{A}$ is less than 0.1 in
the Galactic
environments. Given an injection scale of turbulence at 30 pc, the coherence
length of the strong turbulence is then $\ga 0.3$ pc, which is still larger than
the Larmor radius of most massive dust ($a=10^{-4}$ cm). In fact, the observations
indicate that the correlation length of magnetic field is a few parsecs \citep{spangler96}.
}

\subsubsection{Gyroresonance}\label{subsubsec:gyro}

 {Gyroresonance further accelerates charged grains.
Grains obtain energy by resonant interactions with the
waves if the resonance condition,
$\omega -k_\parallel v\mu =n\Omega$
($n=0,\,\pm 1,\,\pm 2,\,\cdots$), is satisfied, where
$\omega$ is the wave frequency, $k_\parallel$ is the parallel
component of wavevector along the magnetic
field, $v$ is the particle velocity, $\mu$ is the
cosine of the pitch angle relative to the magnetic
field, and $\Omega =qB/(mc)$ is the Larmor frequency of
the particle ($q$ is the charge, $B$ is the magnetic field
strength, $m$ is the grain mass, and $c$ is the
light speed). The above condition indicates that
gyroresonance occurs when the Doppler-shifted frequency
of the wave in the grain's guiding centre rest frame
is a multiple of the gyrofrequency and when the rotating
direction of the electric wavevector is the same as the
direction of the Larmor gyration of the grain. Then
the steady state distribution function of grains
is calculated by a Fokker-Planck equation treating the
effects of gyroresonance acceleration and gaseous
friction.}

 {Gyroresonance is efficient for large grains: The
condition for gyroresonance is that
the Larmor frequency $\Omega$ is smaller than the
the cutoff wave frequency of the turbulence $\omega_\mathrm{c}$.
The velocity of accelerated grain only weakly depends on the
charge and mass as long as the aforementioned condition is
satisfied (YLD04). On the other hand, the energy gain rate of
the grains scales linearly with the intensity of MHD turbulence;
thus the velocity is roughly proportional to the square root of
the intensity of the MHD turbulence.} 

\subsubsection{Overall features of grain velocity}

 {The results in YLD04 indicate that gyroresonance
accelerates silicate with $a\ga 2\times 10^{-5}$ cm and
graphite with $a\ga 3\times 10^{-5}$ cm to velocities large
enough ($\sim 20$ km s$^{-1}$) for shattering in WNM.
Both silicate and graphite grains with
$a\ga\mbox{several}\times 10^{-6}$ cm achieve
velocities (1--2 km s$^{-1}$) near to the shattering
thresholds in CNM (Table \ref{tab:species}).
Although silicate and graphite with
$a\ga 10^{-6}$ cm are accelerated to 20 km s$^{-1}$
by gyroresonance in WIM, the acceleration by hydrodrag
is larger because the dissipation of
turbulence in WIM is less than that in WNM.}

\begin{table*}
\centering
\begin{minipage}{150mm}
\caption{Summary of grain properties.}
\begin{tabular}{@{}lccccccccc@{}}\hline
Species & $\rho_\mathrm{gr}$ & $c_0$
& $s$ & $v_\mathrm{shat}$ & $P_1$ & $P_v$ & $\gamma$ &
$E$ & $\nu$ \\
& (g cm$^{-3}$) & (km s$^{-1}$) & & (km s$^{-1}$) & (dyn cm$^{-2}$)
& (dyn cm$^{-2}$) & (erg cm$^{-2}$) & (dyn cm$^{-2}$) &
 \\ \hline
Silicate & 3.3 & 5   & 1.2 & 2.7 & $3\times 10^{11}$ &
$5.4\times 10^{12}$ & 25 & $5.4\times 10^{11}$
& 0.17 \\
Graphite & 2.2 & 1.8 & 1.9 & 1.2 & $4\times 10^{10}$ &
$5.8\times 10^{12}$ & 12 & $3.4\times 10^{10}$
& 0.5 \\
\hline
\end{tabular}
\label{tab:species}
\end{minipage}
\end{table*}

On the other hand, low relative velocities of small
grains allow coagulation to occur. Moreover, with small
velocities, a dense environment is necessary for a high
enough collision rate. Thus, coagulation is important in
DC for
grains with $a\la 10^{-6}$ cm, which have velocities
$\la 10^3$ cm s$^{-1}$. Coagulation is also possible in
MC with a smaller rate.

Since YLD04 only calculated the grain velocity for
$a\geq 10^{-6}$~cm except for WIM, we extend the
calculations down to $a=10^{-7}$ cm. Below
$a\sim 10^{-6}$ cm, however, the coupling between gas
and grains occurs on a scale smaller than the
dissipation scale of turbulence.
Thus, the velocities of grains typically smaller
than $10^{-6}$ cm are determined by the thermal velocities.

\subsection{Time evolution of the grain size distribution}
\label{subsec:timeev}

We assume that grains are spherical with a constant
material density $\rho_\mathrm{gr}$. The mass $m$ and the
radius $a$ of a grain are related by
\begin{eqnarray}
m=\frac{4\pi}{3}a^3\rho_\mathrm{gr}\, .
\end{eqnarray}
The number density of grains whose radii are between $a$
and $a+\mathrm{d}a$ is denoted as $n(a)\,\mathrm{d}a$,
where the entire range of $a$ is from $a_\mathrm{min}$ to
$a_\mathrm{max}$. The total grain mass is conserved in
shattering and coagulation. To ensure the conservation of
the total mass of grains, it is numerically convenient to
consider the distribution function of grain mass instead
of grain size. We denote the number density of grains
whose masses are
between $m$ and $m+\mathrm{d}m$ as $\tilde{n}(m)\,\mathrm{d}m$.
The two distribution functions are related as
$n(a)\,\mathrm{d}a=\tilde{n}(m)\,\mathrm{d}m$.

For numerical calculation, we consider $N$ discrete bins for
the grain radius. The grain radius in the $i$-th
($i=1,\,\cdots ,\, N$) bin is between $a_{i-1}^\mathrm{(b)}$
and $a_i^\mathrm{(b)}$, where
$a_{i}^\mathrm{(b)}=a_{i-1}^\mathrm{(b)}\delta$,
$a_0^\mathrm{(b)}=a_\mathrm{min}$, and
$a_N^\mathrm{(b)}=a_\mathrm{max}$ (i.e.\
$\log\delta$ specifies the width of a logarithmic bin:
$\log\delta =(1/N)\log (a_\mathrm{max}/a_\mathrm{min})$).
We represent the grain radius and mass in the $i$-th bin
with
$a_i\equiv (a_{i-1}^\mathrm{(b)}+a_i^\mathrm{(b)})/2$ and
$m_i\equiv (4\pi /3)a_i^3\rho_\mathrm{gr}$. The boundary
of the mass bin is defined as
$m_i^\mathrm{(b)}\equiv (4\pi /3)[a_i^\mathrm{(b)}]^3
\rho_\mathrm{gr}$.
Giving $a_\mathrm{min}$, $a_\mathrm{max}$, and $N$,
all bins can be set. A grain in the $i$-th bin
is called ``grain $i$''.
In this paper we take $N=32$ after confirming
that the results do not change if we take a larger
$N$. For the size range, we assume
$a_\mathrm{min}=0.001~\mu$m and $a_\mathrm{max}=0.25~\mu$m
to reproduce the Milky Way extinction curve
(Section \ref{subsec:initial}).

The mass density of grains contained in the $i$-th bin,
$\tilde{\rho}_i$, is defined as
\begin{eqnarray}
\tilde{\rho}_i\equiv m_i\tilde{n}(m_i)(m_i^\mathrm{(b)}-
m_{i-1}^\mathrm{(b)})\, .
\end{eqnarray}
Then, the time evolution of $\tilde{\rho}_i$ is expressed
as
\begin{eqnarray}
\frac{\mathrm{d}\tilde{\rho}_i}{\mathrm{d}t}=
\left[\frac{\mathrm{d}\tilde{\rho}_i}{\mathrm{d}t}
\right]_\mathrm{shat}+
\left[\frac{\mathrm{d}\tilde{\rho}_i}{\mathrm{d}t}
\right]_\mathrm{coag}
\, ,
\end{eqnarray}
where the first and the second terms in the right-hand
side are the contributions from shattering and
coagulation, respectively. These two terms are estimated
in Sections \ref{subsubsec:shat} and \ref{subsubsec:coag}.

We consider silicate and graphite as grain species. In
order to avoid complexity in compound species, we only
treat collisions between the same species. Although this
underestimates the grain collision rate by a factor of
$\sim 2$, our simple assumption here is enough to
understand the effects of interstellar turbulence on the
grain size distribution for the first time.
The adopted parameters for each grain species
are summarized in Table \ref{tab:species}
and are taken from JTH96 and \citet{chokshi93}.
We use the same notation ($n(a)$) for both silicate
and graphite size distributions.

\subsubsection{Shattering}\label{subsubsec:shat}

The time evolution of $\tilde{\rho}_i$ by shattering can be
written as
\begin{eqnarray}
\left[\frac{\mathrm{d}\tilde{\rho}_i}{\mathrm{d}t}
\right]_\mathrm{shat}\hspace{-2mm}
& =\hspace{-2mm} &
-m_i\tilde{\rho}_i
\sum_{k=1}^{N}\alpha_{ki}\tilde{\rho}_k+
\sum_{j=1}^{N}\sum_{k=1}^N\alpha_{kj}\tilde{\rho}_k
\tilde{\rho}_jm_\mathrm{shat}^{kj}(i)\, , \nonumber\\
& &
\label{eq:time_ev}
\end{eqnarray}
\begin{eqnarray}
\alpha_{ki}=\left\{
\begin{array}{ll}
{\displaystyle \frac{\sigma_{ki}v_{ki}}{m_im_k}} &
\mbox{if $v_{ki}>v_\mathrm{shat}$,} \\
0 & \mbox{otherwise,}
\end{array}
\right.
\end{eqnarray}
where $m_\mathrm{shat}^{kj}(i)$ is the total mass of the
shattered fragments of a grain $k$ that enter the $i$-th
bin in the collision between grains $k$ and $j$,
$\sigma_{ki}$ and $v_{ki}$ are, respectively, the grain-grain
collisional cross section and the relative collision speed
between grains $k$ and $i$, and $v_\mathrm{shat}$ is the
velocity threshold for shattering to occur.
For the cross section, we apply
$\sigma_{ki}=\pi (a_k+a_i)^2$.

The grain velocities given by YLD04 are typical velocity
dispersions. The relative velocity $v_{ki}$ is treated with
a similar manner to Appendix A of JTH96. Each time
step is divided into 4 small steps, and we apply
$v_{ik}=v_i+v_k$, $|v_i-v_k|$, $v_i$, and $v_k$ in each small
step, where $v_i$ and $v_k$
are the velocities of grains $i$ and $k$, respectively
(see Section \ref{subsec:vel}).
Note that the mass distribution of the
shattered fragment $m_\mathrm{shat}^{kj}(i)$ depends on
$v_{kj}$. The method for calculating
$m_\mathrm{shat}^{kj}(i)$ is described in
Section \ref{subsec:frag}.

\subsubsection{Coagulation}\label{subsubsec:coag}

The time evolution of $\tilde{\rho}_i$ by coagulation can
be written in a similar form to equation (\ref{eq:time_ev}):
\begin{eqnarray}
\left[\frac{\mathrm{d}\tilde{\rho}_i}{\mathrm{d}t}
\right]_\mathrm{coag}
& \hspace{-2mm}= & \hspace{-2mm}
-m_i\tilde{\rho}_i
\sum_{k=1}^{N}\alpha_{ki}\tilde{\rho}_k+
\sum_{j=1}^{N}\sum_{k=1}^N\alpha_{kj}\tilde{\rho}_k
\tilde{\rho}_jm_\mathrm{coag}^{kj}(i)\, ,\nonumber\\
& &
\end{eqnarray}
\begin{eqnarray}
\alpha_{ki}=\left\{
\begin{array}{ll}
{\displaystyle \frac{\sigma_{ki}v_{ki}}{m_im_k}} &
\mbox{if $v_{ki}<v_\mathrm{coag}^{ki}$,} \\
0 & \mbox{otherwise.}
\end{array}
\right.
\end{eqnarray}
Here, $m_\mathrm{coag}^{kj}(i)=m_i$ if
$m_{i-1}^\mathrm{(b)}\leq m_k+m_j<m_i^\mathrm{(b)}$; otherwise
$m_\mathrm{coag}^{kj}(i)=0$. The coagulation is assumed to occur
only if the relative velocity is less than the coagulation
threshold velocity $v_\mathrm{coag}^{ki}$. The coagulation
threshold velocity is given by
(\citealt{chokshi93,dominik97}; YLD04)
\begin{eqnarray}
v_\mathrm{coag}^{ki}=2.14F_\mathrm{stick}\left[
\frac{a_k^3+a_i^3}{(a_k+a_i)^3}\right]^{1/2}
\frac{\gamma^{5/6}}{E^{1/3}R_{ki}^{5/6}\rho_\mathrm{gr}^{1/2}}\, ,
\label{eq:vcoag}
\end{eqnarray}
where a factor $F_\mathrm{stick}=10$ is introduced following
YLD04 (based on the experimental work by \citealt{blum00}),
$\gamma$ is the surface energy per unit area,
$R_{ki}\equiv a_ka_i/(a_k+a_i)$ is the reduced radius of the
grains, $E$ is related to Poisson's ratios ($\nu_k$ and
$\nu_i$) and Young's modulus ($E_k$ and $E_i$) by
$1/E\equiv (1-\nu_k)^2/E_k+(1-\nu_i)^2/E_i$. The values of
$\gamma$ and $E$ are taken from Table 3 of \citet{chokshi93}
(the data for
quartz and graphite are used for silicate and graphite,
respectively) as summarized in Table \ref{tab:species}.
The treatment of $v_{ki}$ in coagulation is the same as that
in shattering (Section \ref{subsubsec:shat}).

 {Although the above form of coagulation threshold velocity
is derived based on both physical and experimental basis, there
could be significant uncertainties. A change of the coagulation
threshold affects the largest size of grains subject to
coagulation, since
larger grains have larger velocities. For example, as shown
later, the coagulation condition is satisfied for grains with
$a\la\mbox{a few}\times 10^{-6}$ cm. However, we have
confirmed that even if
the coagulation threshold is altered by a factor of ten from
the above values, the size of grains subject to
coagulation changes only by a factor of 2--3. This is
because of a steep dependence of the grain velocity on
the grain size.
}

\subsection{Production of shattered fragments}
\label{subsec:frag}

Here we determine the mass distribution of the shattered
fragments, $m_\mathrm{shat}^{kj}(i)$, in
equation (\ref{eq:time_ev}).
 {The shattering rate is determined mostly by the
collision frequency between grains. Thus, the overall
results in this paper is less sensitive to the
detailed model of shattered fragments than to the
grain velocities. Indeed, JTH96 show that the size
distribution of shattered fragments does not have a large
influence on the overall grain size distribution.
We have also confirmed that the specific parameters
adopted in this subsection do not affect the results
significantly. We can have an idea about the uncertainties
caused by the material
parameters by comparing the results for silicate and graphite.
Since the results in these two materials are broadly
similar,\footnote{If the results are very different
between silicate and graphite, the difference can be
attributed to the difference in grain velocity.} the
assumption on the material parameters only has a secondary
importance in our models as long as there are not very
eccentric materials involved.
}

\begin{figure}
\begin{center}
\includegraphics[width=8cm]{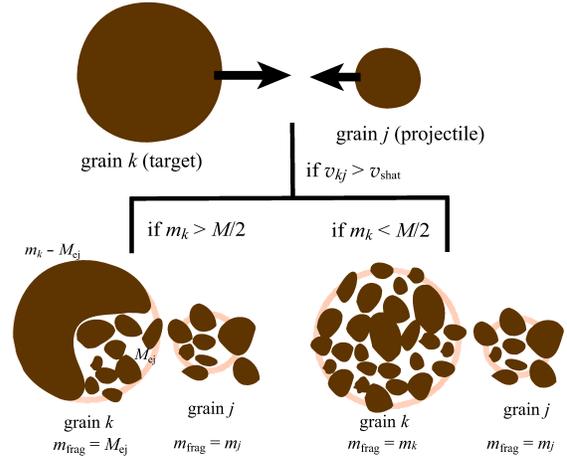}
\end{center}
\caption{Schematic figure of our treatment of shattering.
We consider a collision of two grains in the $k$-th
size bin and the $j$-th size bin (called grain $k$ and
grain $j$, respectively) with a relative velocity of
$v_{kj}$. We call the larger and smaller grains ``target''
and ``projectile'', respectively. Here we assume $k\geq j$;
that is, the target is grain $k$ and the projectile is grain
$j$. If the shocked mass $M$
in the target is larger than the mass of the target
$m_k$ ($M>m_k/2$), we assume that the entire target
fragments into small pieces ($m_\mathrm{frag}=m_k$). If
$M<m_k/2$, a part of the target mass, $M_\mathrm{ej}$ is
shattered and ejected ($m_\mathrm{frag}=M_\mathrm{ej}$). The
entire projectile is assumed to fragment ($m_\mathrm{frag}=m_j$).
The size distribution of the fragments is assumed to follow
equation (\ref{eq:frag}).
\label{fig:schematic}}
\end{figure}

We illustrate our treatment of shattering in
Fig.\ \ref{fig:schematic}. We consider a
collision between grains $k$ and $j$ (here we assume
$k\geq j$), and call the grains labeled as $k$ and $j$
target and
projectile, respectively. The necessary material quantities
are summarized in Table~\ref{tab:species}.
The mass shocked to the critical pressure
for cratering in the target, $M$, is given by
(\citealt{tielens94};
JTH96)\footnote{In JTH96, $\sigma_r$ is denoted as
$\sigma_{1i}$.}
\begin{eqnarray}
\frac{M}{m_j}=\frac{1+2{\cal R}}{(1+{\cal R})^{9/16}}
\frac{1}{\sigma_{r}^{1/9}}\left(
\frac{{\cal M}_r^2}{\sigma_1{\cal M}_1^2}\right)^{8/9}\,
,
\end{eqnarray}
where ${\cal R}=1$ in the collision between the
same species (we adopt ${\cal R}=1$ in this paper),
${\cal M}_r\equiv v_{kj}/c_0$ ($c_0$ is the sound
speed of the grain material), $\sigma_1$ and $\sigma_{1i}$
are constants typically of order unity
(equation \ref{eq:sigma}), and ${\cal M}_1$ is
the Mach number corresponding to the critical
pressure $P_1$:
\begin{eqnarray}
{\cal M}_1=\frac{2\phi_1}{1+(1+4{s}\phi_1)^{1/2}}\, ,
\end{eqnarray}
where $\phi_1\equiv P_1/(\rho_\mathrm{gr}c_0^2)$, and
${s}$ is a dimensionless material constant that
determines the relation between the shock velocity and
the velocity of the shocked matter.
Using the following expression for $\sigma$ as
\begin{eqnarray}
\sigma ({\cal M})\equiv
\frac{0.30({s}+{\cal M}^{-1}-0.11)^{0.13}}
{{s}+{\cal M}^{-1}-1}\, ,\label{eq:sigma}
\end{eqnarray}
we obtain $\sigma_{1}=\sigma ({\cal M}_1)$ and
$\sigma_{1i}=\sigma ({\cal M}_r/(1+{\cal R}))$.
We assume that if more than the half of the target is shocked
(i.e.\ $M>m_k/2$) the entire target is shattered
(i.e.\ $m_\mathrm{frag}=m_k$ in equation \ref{eq:norm},
where $m_\mathrm{frag}$ is the total mass of the fragments).
Otherwise, only a fraction of the target mass
($M_\mathrm{ej}$) is ejected from the target
(i.e.\ $m_\mathrm{frag}=M_\mathrm{ej}$).
$M_\mathrm{ej}$ is assumed to be $0.40M$, i.e.\ 40\% of the
shocked mass is finally ejected. This fraction is
derived for $z=3.4$, where the radial velocity of the cratering
flow in the shattered material is approximated to be
$\propto R^{-z}$ ($R$ is the distance from the cratering centre;
JTH96).
Finally, the entire projectile is assumed to fragment
into small pieces (i.e.\ $m_\mathrm{frag}=m_j$ for all projectiles).

The fragments are assumed to follow the size
distribution (\citealt{hellyer70}; JTH96)
\begin{eqnarray}
n_\mathrm{frag}(a)\,\mathrm{d}a=C_\mathrm{frag}a^{-3.3}\,
\mathrm{d}a\, ,\label{eq:frag}
\end{eqnarray}
where the normalization constant $C_\mathrm{frag}$
is determined by
\begin{eqnarray}
m_\mathrm{frag}=\int_{a_\mathrm{fmin}}^{a_\mathrm{fmax}}
mn_\mathrm{frag}(a)\,
\mathrm{d}a\, .\label{eq:norm}
\end{eqnarray}
Here, $a_\mathrm{fmin}$ and $a_\mathrm{fmax}$,
respectively, specify the upper and lower bounds
of the fragment radius, which are determined in
Sections \ref{subsubsec:projectile} and
\ref{subsubsec:target} for the projectile and the target,
respectively. If $a_\mathrm{fmin}$ estimated is less than
$a_\mathrm{min}$, we take $a_\mathrm{fmin}=a_\mathrm{min}$.
If $a_\mathrm{fmax}$ is also smaller than $a_\mathrm{min}$,
all the fragments are put in the bin with the
smallest size (i.e.\ $i=0$). Finally, the mass of
shattered fragments
in the $i$-th bin is determined in terms of
$n_\mathrm{frag}$ as
\begin{eqnarray}
m_\mathrm{shat}(i)=\int_{a_{i-1}^\mathrm{(b)}}^{a_i^\mathrm{(b)}}
mn_\mathrm{frag}(a)\,\mathrm{d}a\, ,
\end{eqnarray}
where the integration is performed in the range
corresponding to the $i$-th bin. We put this mass
in equation (\ref{eq:time_ev}) (note that the superscript
$kj$ is omitted here).

In the following, we summarize how to determine
$a_\mathrm{fmin}$ and $a_\mathrm{fmax}$.

\subsubsection{Projectile $m_j$}
\label{subsubsec:projectile}

The entire projectile is assumed to fragment into
small pieces, i.e.\ $m_\mathrm{frag}=m_j$.
The maximum  grain size, $a_\mathrm{fmax}$, is
determined by (JTH96)
\begin{eqnarray}
a_\mathrm{fmax}=0.22a_j\left(\frac{v_\mathrm{cat}}{v_{kj}}
\right)\, ,\label{eq:famax}
\end{eqnarray}
where $v_\mathrm{cat}$ is the critical spalling
collision velocity given by
\begin{eqnarray}
v_\mathrm{cat}=c_0\left[\frac{m_k}{(1+2{\cal R})m_j}
\right]^{9/16}\sigma_1^{1/2}\sigma_{1i}^{1/16}
(1+{\cal R}){\cal M}_1\, .
\end{eqnarray}
The minimum grain size, $a_\mathrm{fmin}$, is determined
by
\begin{eqnarray}
a_\mathrm{fmin}=0.03a_\mathrm{fmax}\, .
\label{eq:famin}
\end{eqnarray}

\subsubsection{Target $m_k$}\label{subsubsec:target}

If $M/m_k>0.5$, we assume that the entire target
fragments into small pieces; i.e.\ $m_\mathrm{frag}=m_k$.
The maximum and minimum fragment sizes are determined
by equations (\ref{eq:famax}) and (\ref{eq:famin}),
respectively, but $j$ and $k$ are exchanged.

If $M/m_k\leq 0.5$, we assume that the mass
$M_\mathrm{ej}$ ($=0.40M$) fragments into smaller grains
(i.e.\ $m_\mathrm{frag}=M_\mathrm{ej}$), and a grain
with mass of $m_k-M_\mathrm{ej}$ is left, which is put in
the corresponding bin. According to equations (10) and
(15) in JTH96, the largest fragment size and the total ejected
volume ($M_\mathrm{ej}/\rho_\mathrm{gr}$) can be related by
\begin{eqnarray}
M_\mathrm{ej}/\rho_\mathrm{gr}=\frac{16}{3}\pi
\frac{z^3(z-2)}{z+1}a_\mathrm{fmax}^3\, .
\end{eqnarray}
We determine $a_\mathrm{fmax}$ according to this equation
with $z=3.4$. The following estimate is adopted for
$a_\mathrm{fmin}$ (JTH96):
\begin{eqnarray}
a_\mathrm{fmin}=a_\mathrm{fmax}\left(\frac{P_1}{P_v}
\right)^{1.47}\, ,
\end{eqnarray}
where $P_v$ is the critical pressure for vaporization
(Table \ref{tab:species}).

\subsection{Initial Grain Size Distribution}
\label{subsec:initial}

It is not an easy task to select a good initial grain
size distribution, since the grain production in stellar
mass loss is not fully understood yet.
Thus, we concentrate on how the standard grain
size distribution is modified by shattering and
coagulation in various ISM phases. As the standard
grain size distribution, we adopt
\begin{eqnarray}
n(a)={\cal C}a^{-K}~(a_\mathrm{min}\leq a\leq
a_\mathrm{max})\, ,
\end{eqnarray}
where ${\cal C}$ is the normalizing constant. We
select $K=3.5$ as derived by MRN to explain the
observed Milky Way extinction curve.
For the size range, we assume
$a_\mathrm{min}=0.001~\mu$m and $a_\mathrm{max}=0.25~\mu$m
for both graphite and silicate (MRN;
\citealt{li01}), although \citet{li01}
adopt more elaborate functional form (see also
\citealt*{kim94}). In fact,
as shown later, the predicted extinction curve is
broadly consistent with the observed extinction curve
(Section \ref{subsec:ext}). Thus, the
above simple assumption for the size distribution
is enough for our purpose.

The normalization factor ${\cal C}$ is determined
according to the mass density of the grains in the
ISM:
\begin{eqnarray}
{\cal R}m_\mathrm{H}n_\mathrm{H}=
\int_{a_\mathrm{min}}^{a_\mathrm{max}}\frac{4\pi}{3}a^3
\rho_\mathrm{gr}{\cal C}a^{-K}\,\mathrm{d}a\, ,
\label{eq:norm_grain}
\end{eqnarray}
where $n_\mathrm{H}$ is the hydrogen number density given
for each ISM phase in YLD04 (see also Table \ref{tab:yan}),
$m_\mathrm{H}$ is the hydrogen atom
mass, and ${\cal R}$ is the dust-to-hydrogen mass ratio
(i.e.\ dust abundance relative to hydrogen) in the ISM.
We adopt ${\cal R}=4.0\times 10^{-3}$ and
$3.4\times 10^{-3}$ for silicate and graphite,
respectively \citep{takagi03}.
As shown in Section \ref{subsec:ext}, the size distribution
assumed here reproduces the observed Milky Way
extinction curve.

\subsection{Timescales}\label{subsec:timescale}

We calculate the change of grain size distribution in
various ISM phases on
typical timescales. The typical timescale of the phase
change among WIM, WNM, and CNM is
$\sim\mbox{a few}\times 10^7$--$10^8$ yr
\citep{ikeuchi88,odonnell97,hirashita01}. 
For WIM, there is another relevant timescale,
that is, recombination timescale. With hydrogen
number density $\sim 0.1$ cm$^{-3}$ and
temperature $\sim 10^4$ K, the recombination timescale
is roughly $\sim 10^6$ yr \citep{spitzer78}.
As shown later, the grains are shattered too much
in WIM for a time $t\ga 10^7$ yr, so a timescale of
the order of Myr is more appropriate for WIM
(Section \ref{sec:result}).
For denser medium, a short lifetime may be reasonable,
and indeed the lifetime of molecular clouds
is estimated to be $\sim 10^7$ yr
\citep{blitz80,palla02,kawamura07} or shorter
\citep{elmegreen00,hartmann03}.
Thus, we examine
$t<10^7$ yr for MC and DC. These timescales are
also consistent with \citet{odonnell97}.

\subsection{Extinction curves}\label{subsec:ext_curve}

Following \citet{odonnell97}, we use extinction curves
to test our results. We calculate extinction curves by
using the optical constants of astronomical silicate
and graphite taken from \citet{draine84}. Then cross
sections for absorption and scattering are calculated
with Mie theory \citep{bohren83} and weighted for the
grain size distribution. Finally the extinction curves
of silicate and graphite are summed up. The extinction
is normalized to the number density of hydrogen
atoms.

For comparison, the observational data of the standard
interstellar extinction of the Milky Way is taken from
\citet{pei92}. \citet{bohlin78} show that the mean Milky
Way $N_\mathrm{H}/E(B-V)$ ($N_\mathrm{H}$ is the column
density of hydrogen atoms and $E(B-V)$ is the excess
of $B-V$ colour) is $5.8\times 10^{21}$ atoms cm$^{-2}$
mag$^{-1}$. Then by using $A_B=(1+R_V)E(B-V)$
($A_\lambda$ is the extinction in units of magnitude
at wavelength $\lambda$ and $R_V\equiv A_V/E(B-V)$),
and adopting $R_V=3.08$ \citep{pei92}, we obtain
$N_\mathrm{H}/A_B=1.422\times 10^{21}$ atoms cm$^{-2}$
mag$^{-1}$ for the mean Milky Way extinction.
\citet{pei92} lists $\xi (\lambda)\equiv A_\lambda /A_B$
for relevant wavelengths, and the equation
$A_\lambda /N_\mathrm{H}=\xi (\lambda)A_B/N_\mathrm{H}$
can be used to obtain $A_\lambda /N_\mathrm{H}$ for the
mean Milky Way extinction curve. The extinction curves
are often normalized to the
value at $V$ band, but we do not adopt this
normalization, because
the $V$ band extinctions themselves in our models are
significantly affected by a slight change of the size
distribution at
$10^{-6}~\mathrm{cm}<a<10^{-5}~\mathrm{cm}$
(Section \ref{sec:result}).
Since our models are based on a simple analytical
treatment of interstellar turbulence with a single
density, it is not reasonable to adopt a normalization
parameter which is not robust to the change of details in
the models.

It is also known that there is a variation in the
Milky Way extinction curves along various lines of sight.
\citet*{cardelli89} argue that the variation of the
extinction curves can be parametrized by $R_V$.
More recently, \citet{fitzpatrick07} show that the
variance of the extinction curves normalized to
$A_V$ is roughly 20\% at $1/\lambda =8~\mu\mathrm{m}^{-1}$
and roughly 10\% at the 2175\AA\ bump. Although we do not
adopt the normalization at $V$ band in the extinction
curve, these variances provide us with a rough idea as to
how much variation of the extinction curve is
permitted in the Galactic environment.

\section{Results}\label{sec:result}

\subsection{Grain size distribution}\label{subsec:size_result}

\begin{figure*}
\begin{center}
\includegraphics[width=6.5cm]{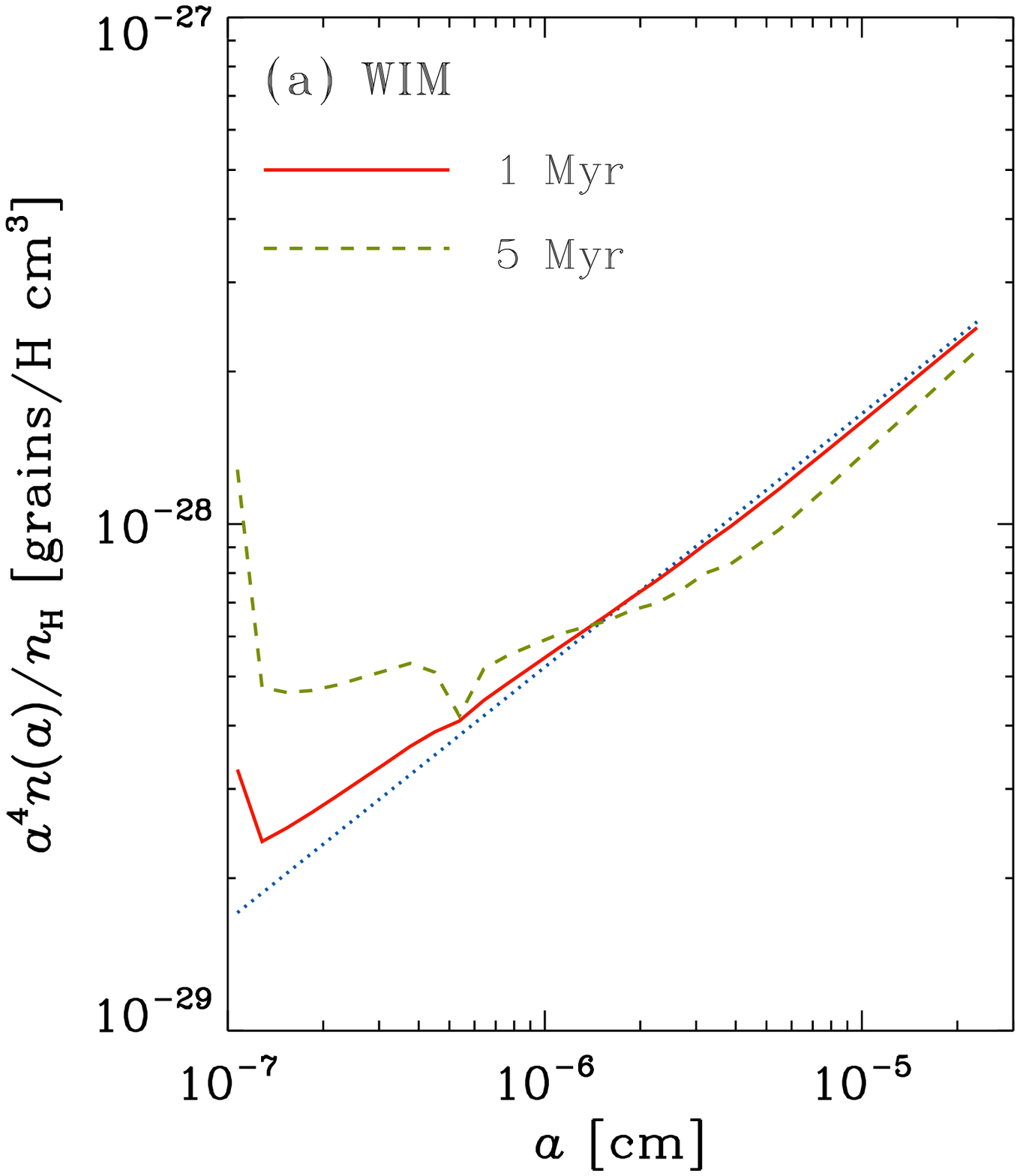}
\includegraphics[width=6.5cm]{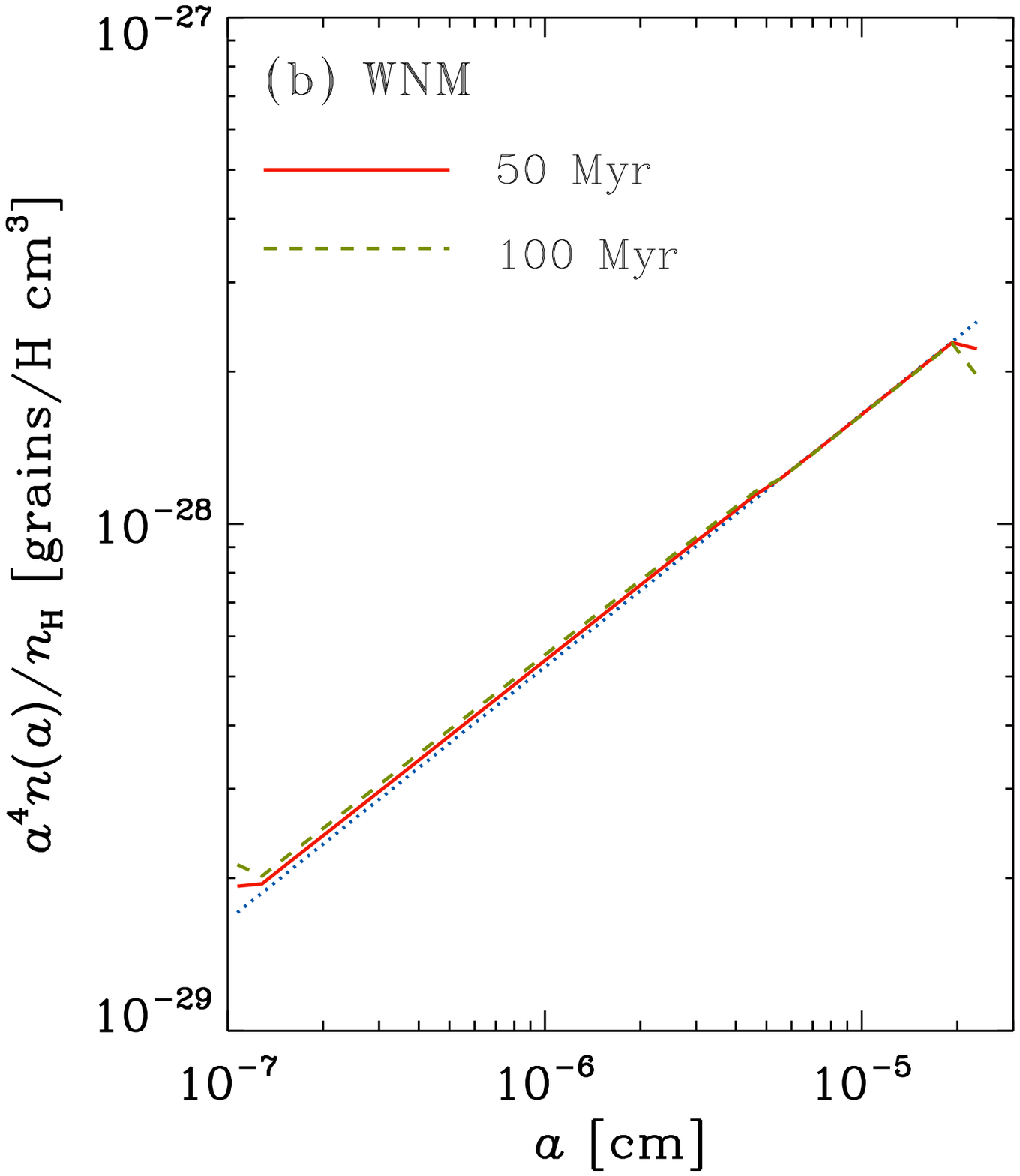}
\includegraphics[width=6.5cm]{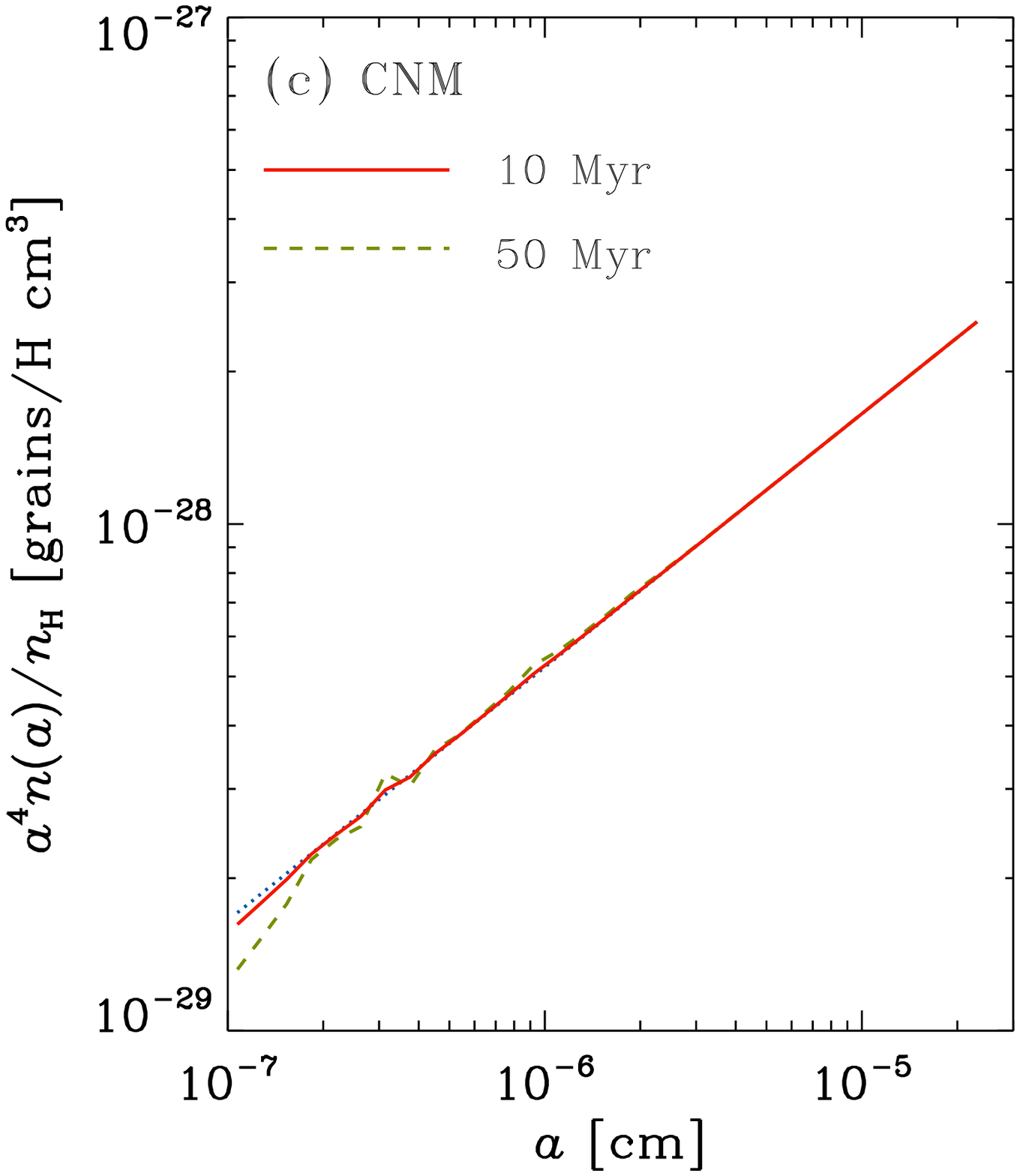}
\includegraphics[width=6.5cm]{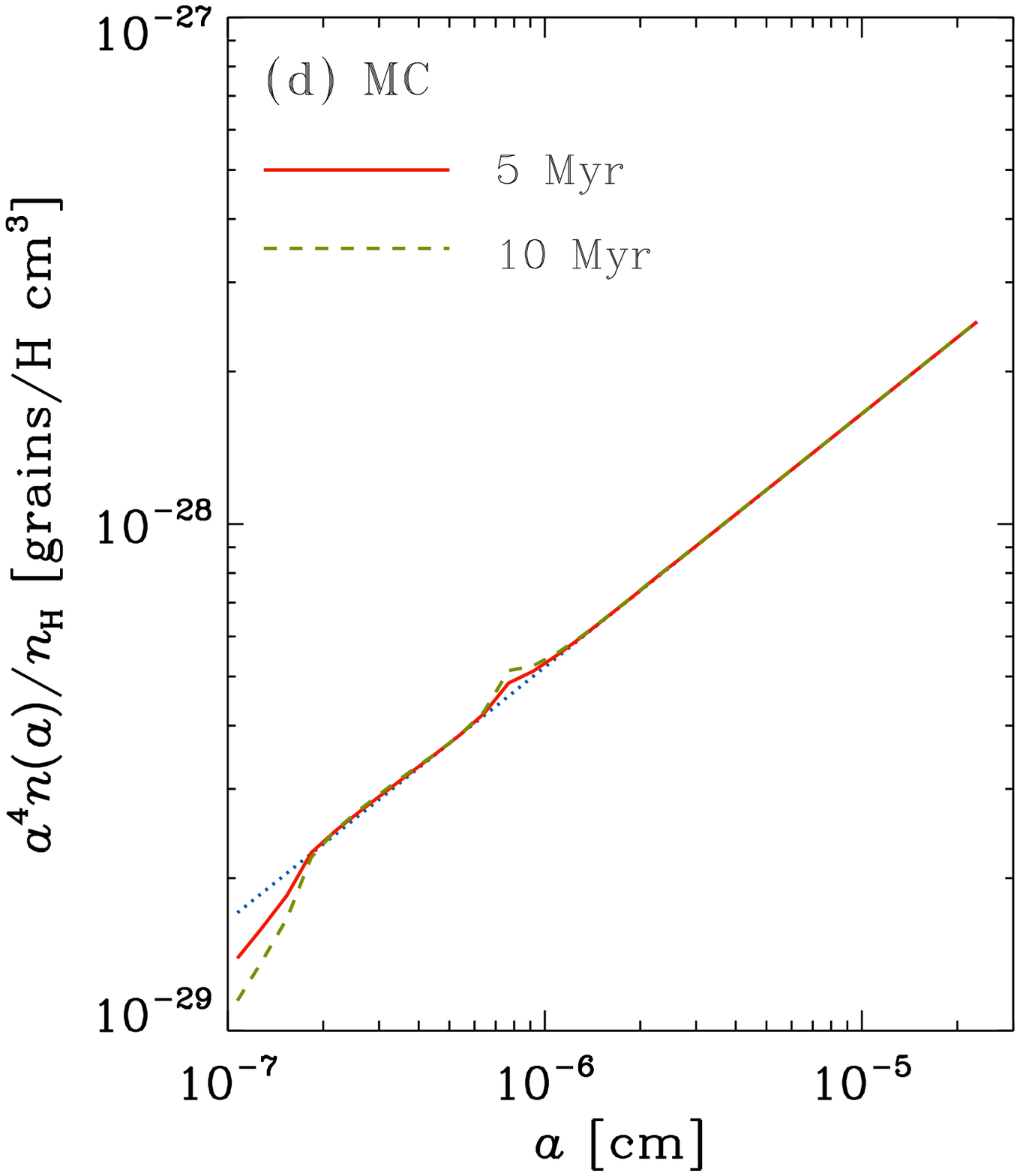}
\includegraphics[width=6.5cm]{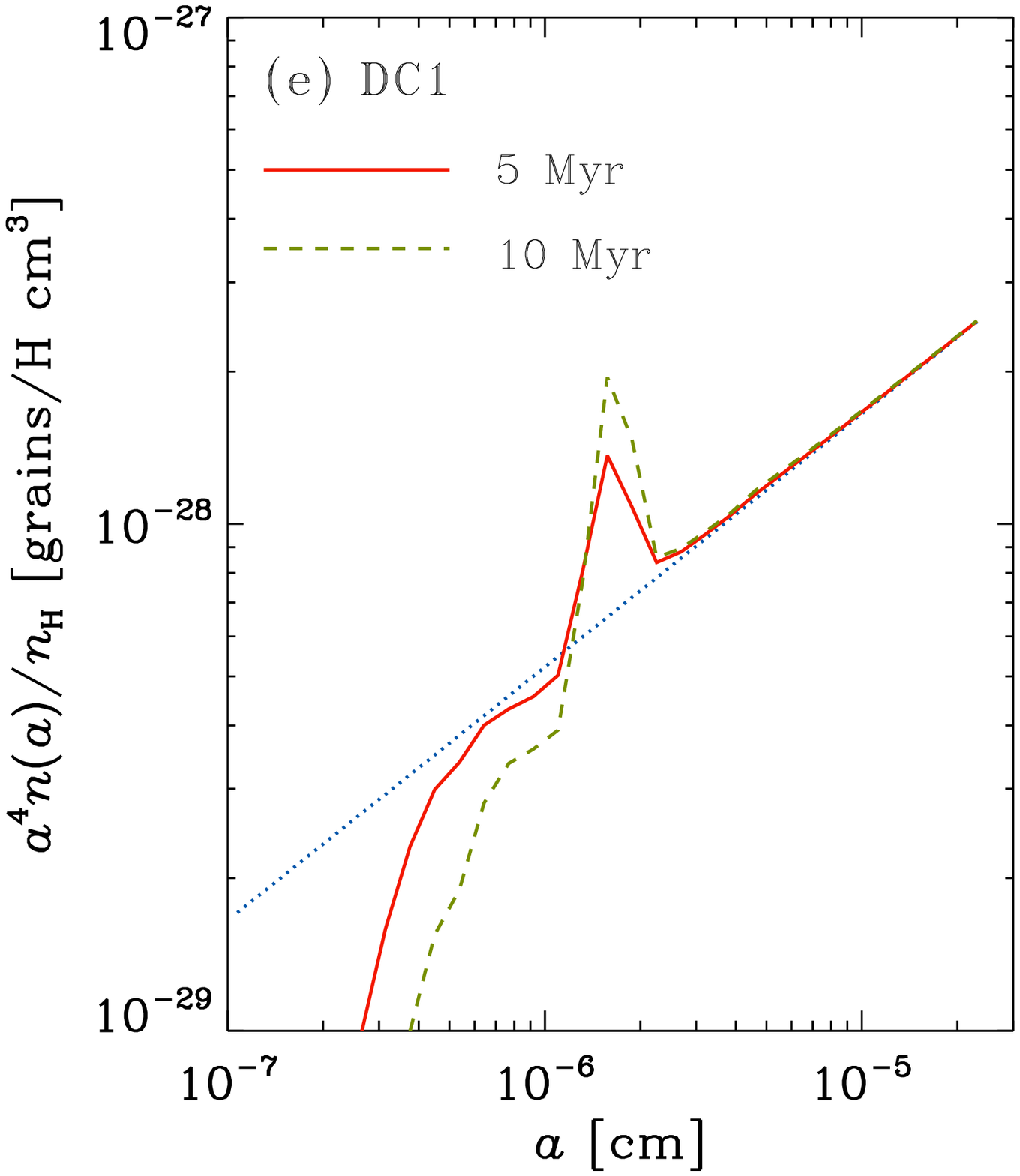}
\includegraphics[width=6.5cm]{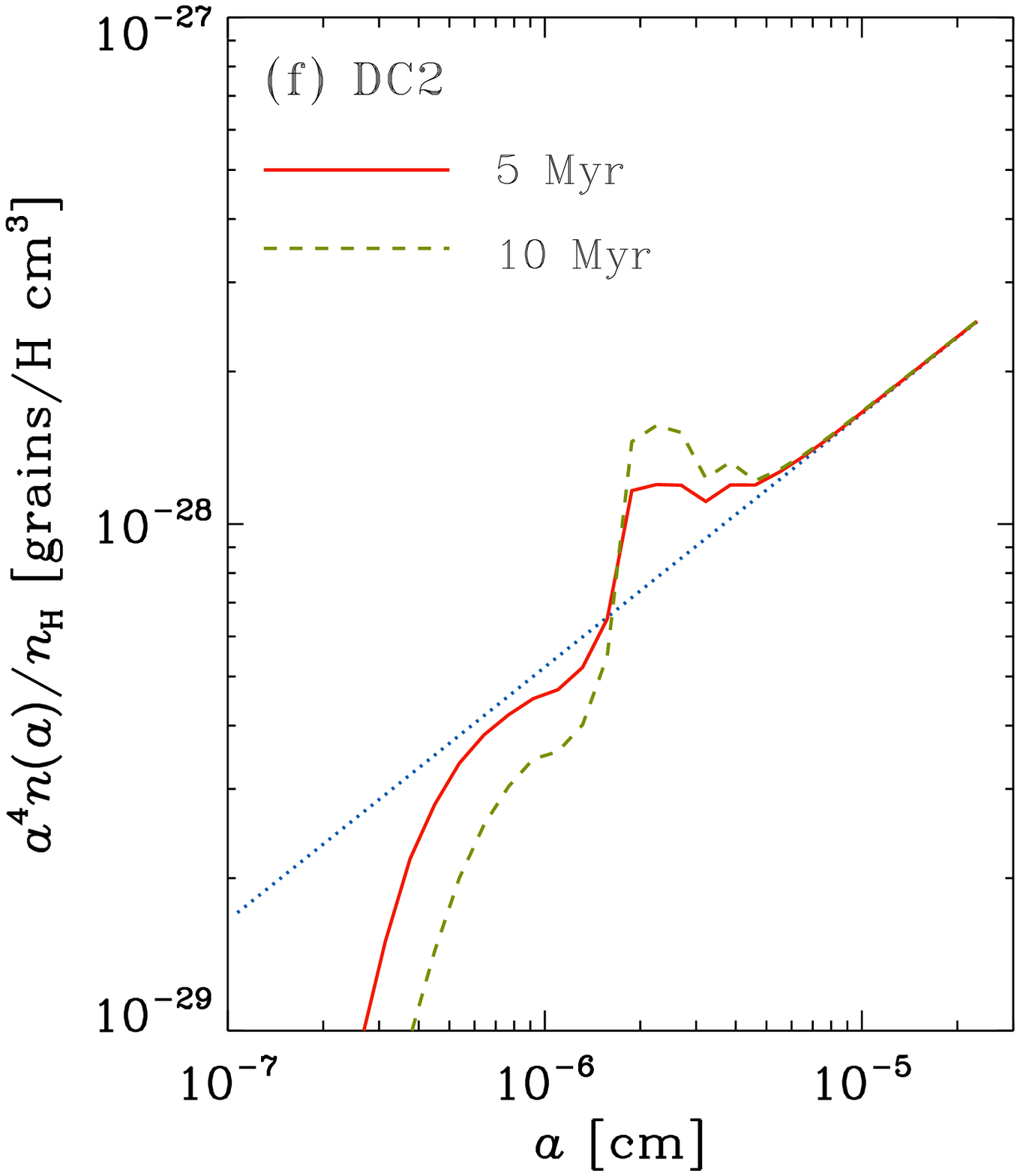}
\end{center}
\caption{The grain size distributions for silicate in
(a) WIM, (b) WNM, (c) CNM, (d) MC, (e) DC1, and (f) DC2.
The initial MRN distribution is shown by the dotted line.
The solid (dashed) line presents the distribution at
$t=1$ Myr ($t=5$ Myr) for Panel (a), $t=10$ Myr and 50 Myr
for Panels (b) and (c), and at $t=5$ Myr ($t=10$ Myr) for
Panels (d), (e), and (f). In this paper, the grain size
distributions are presented by multiplying $a^4$ to show
the mass distribution in each logarithmic bin of the grain
radius. The grain size distribution is normalized to the
hydrogen number density $n_\mathrm{H}$.
\label{fig:sil}}
\end{figure*}

\begin{figure*}
\begin{center}
\includegraphics[width=6.5cm]{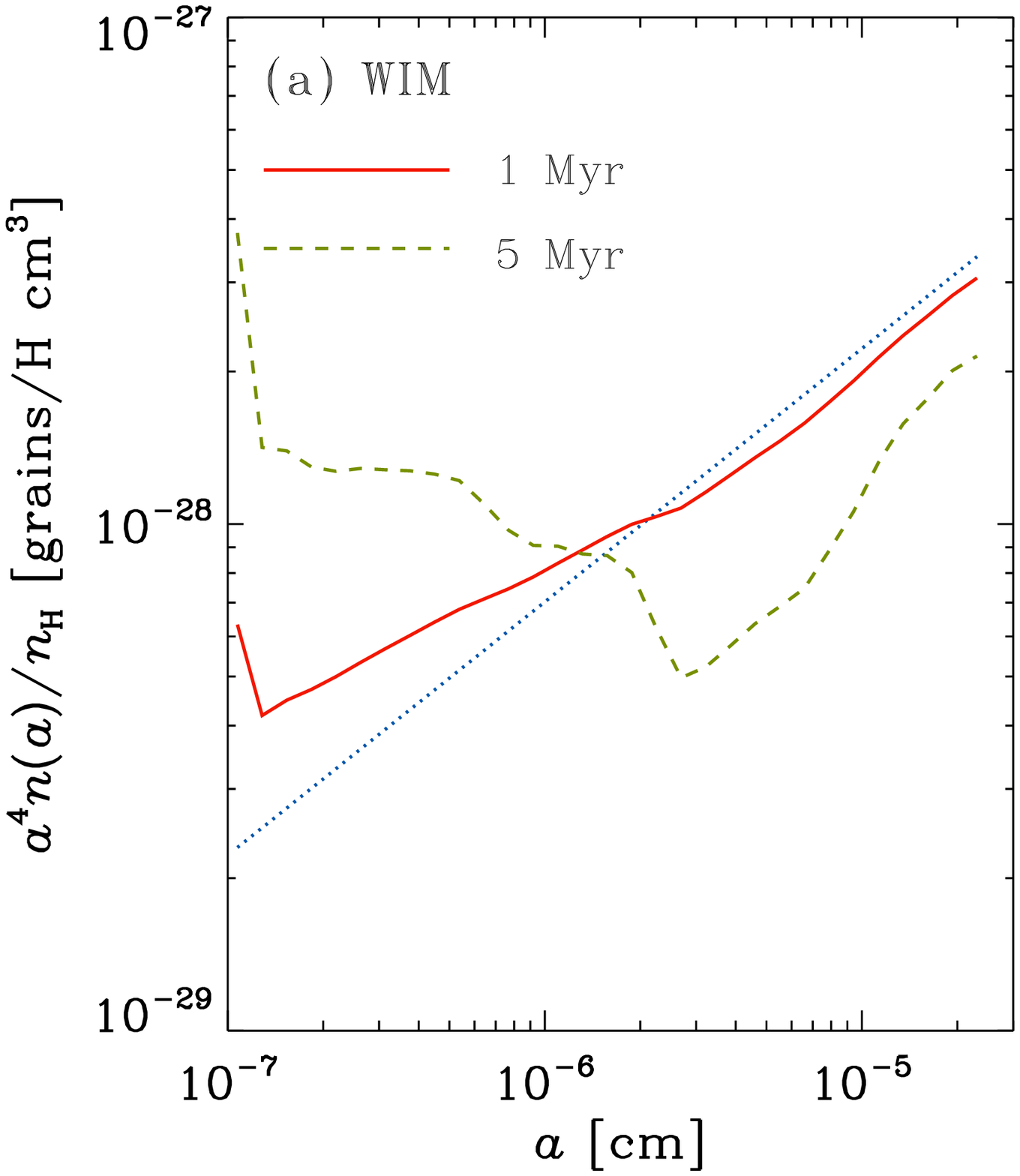}
\includegraphics[width=6.5cm]{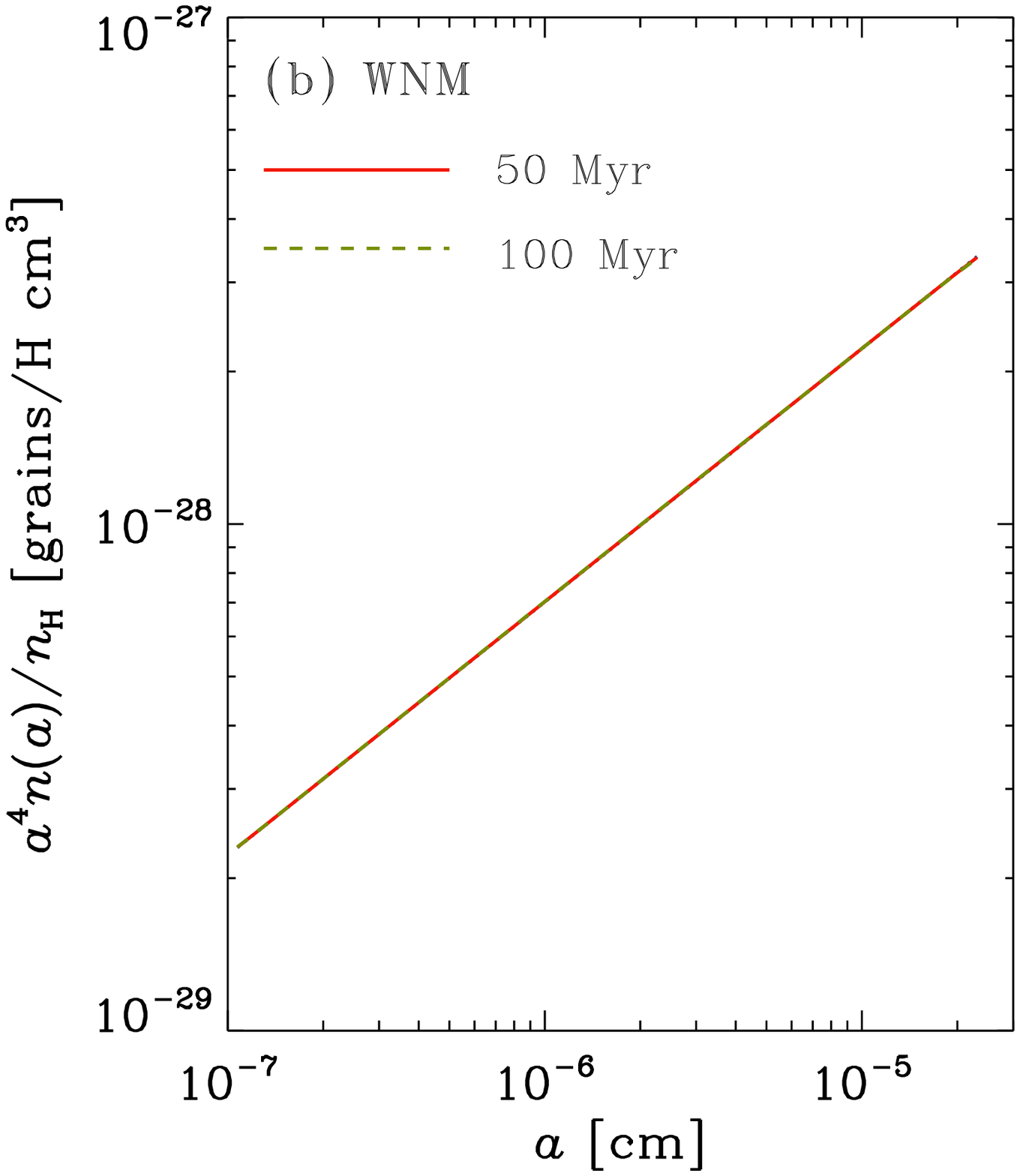}
\includegraphics[width=6.5cm]{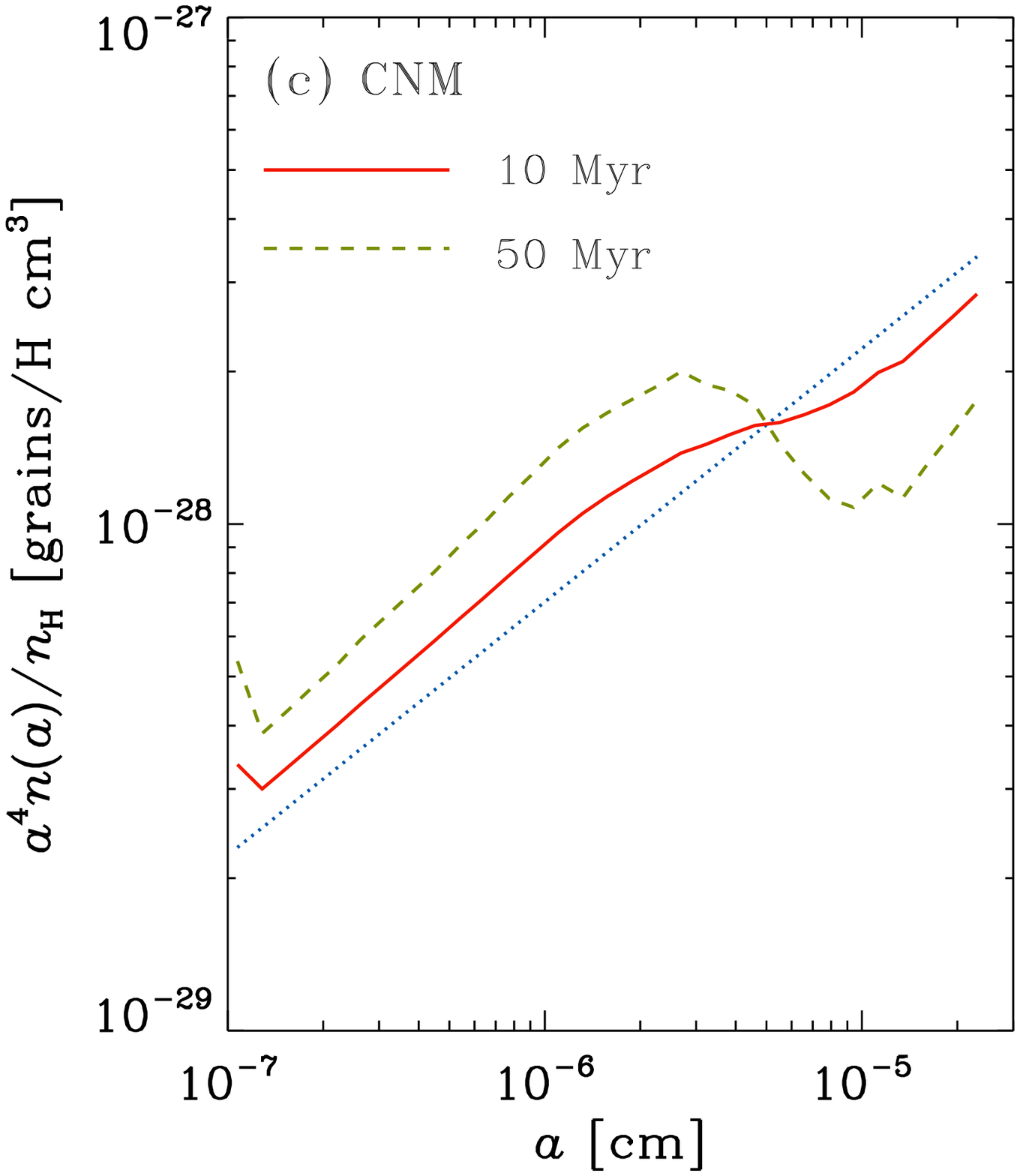}
\includegraphics[width=6.5cm]{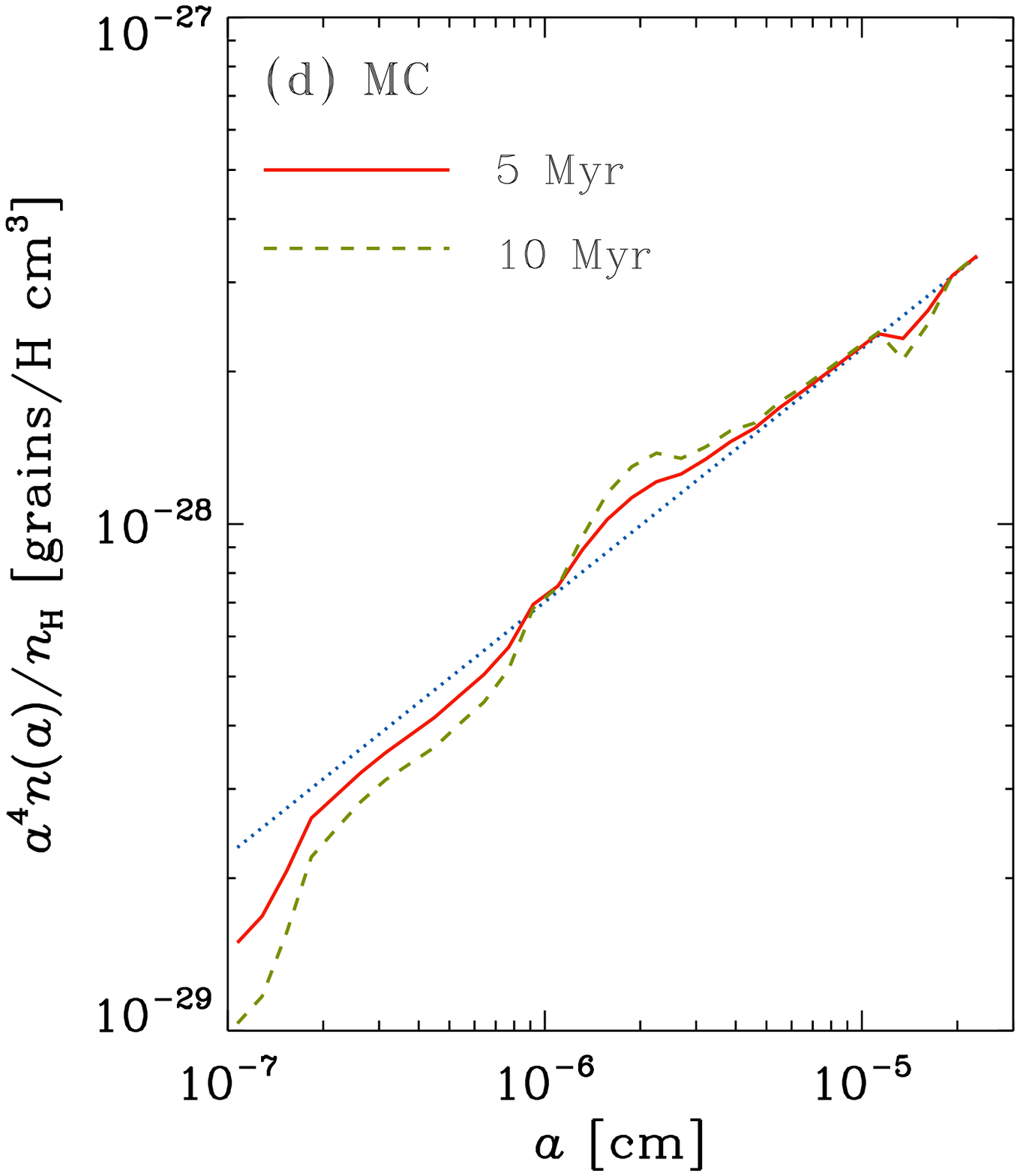}
\includegraphics[width=6.5cm]{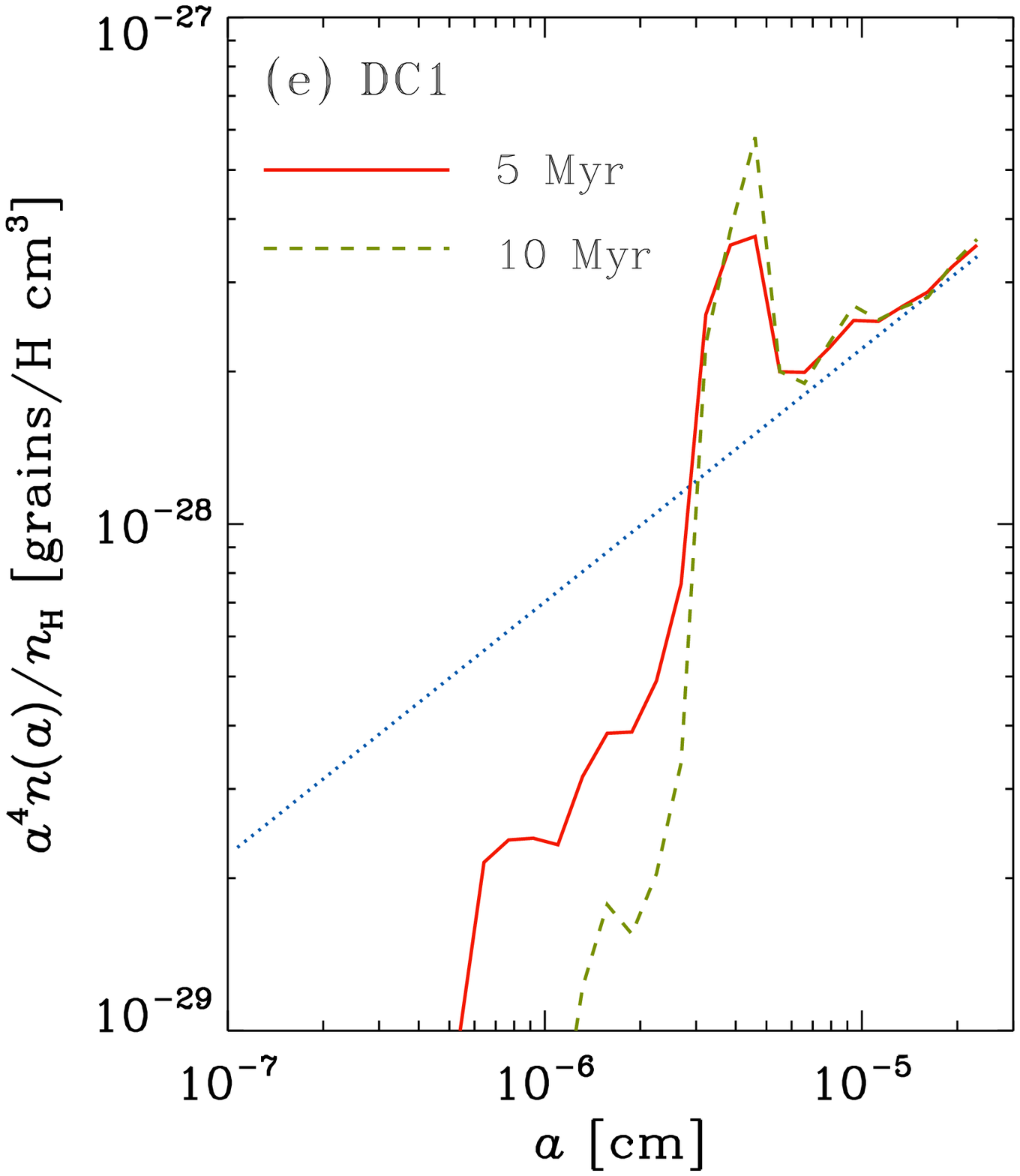}
\includegraphics[width=6.5cm]{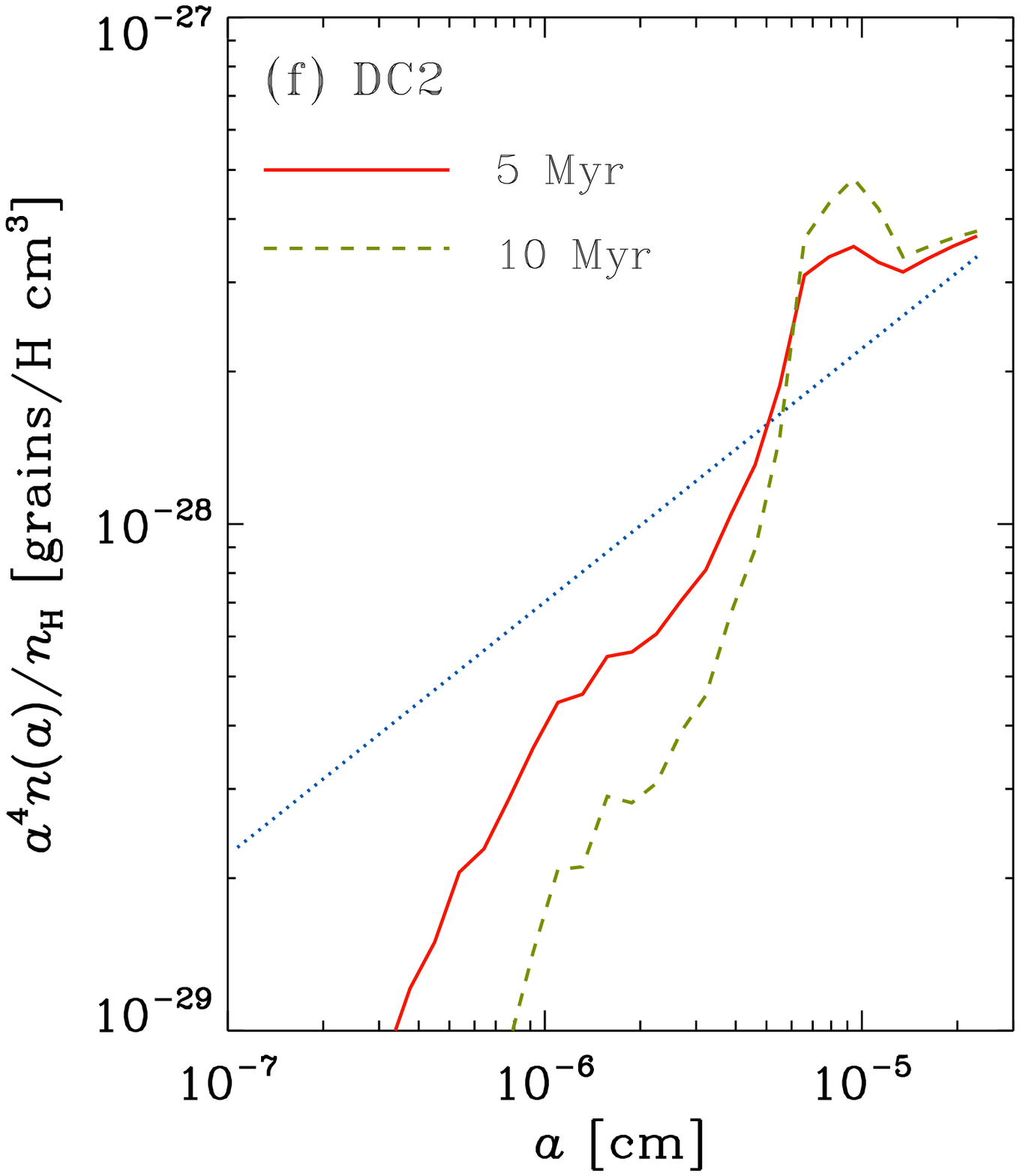}
\end{center}
\caption{Same as Fig.\ \ref{fig:sil} but for
graphite.
\label{fig:gra}}
\end{figure*}

In Figs.\ \ref{fig:sil} and \ref{fig:gra}, we show the
size distributions of silicate and graphite, respectively,
for various ISM phases. The size distribution is expressed
by multiplying $a^4$ to show the ``mass distribution'' in
each logarithmic bin of the grain size \citep{odonnell97};
i.e.\ $a^3$ comes from the grain mass and another factor
$a$ originates from $\mathrm{d}a/\mathrm{d}\ln a=a$. The
largest change is seen in WIM, for which we present the grain
size distributions in the
shortest timescales (1 Myr and 5 Myr). If the grains are
processed for a longer time in WIM, the extinction curves
become too modified to be consistent with the observed
Milky Way extinction curve (Section \ref{subsec:ext}).
In WIM, grains with
$a\ga\mbox{a few}\times 10^{-6}$ cm are efficiently
accelerated by  up to a velocity larger than
the shattering threshold velocity.
If the grain velocity is the same, shattering efficiently
destroys small grains because of their large
surface-to-volume ratios. Thus, the largest shattering
efficiency is realized for the smallest grains which
obtain a velocity above the shattering threshold.
For this reason, grains with $a\sim 10^{-6}$ cm are
the most efficiently shattered in WIM.
This is different from shattering in supernova shocks,
where such small grains are not efficiently shattered
(JTH96) since small grains tend to
be decelerated quickly by the gas drag.

In WNM, because of the ion-neutral collision, the fast
modes are damped on a larger scale than in WIM. Thus,
gyroresonance is not efficient for small grains, and
only large grains with $a>0.2~\mu$m for silicate and
with $a>0.3~\mu$m for graphite can be accelerated
to a velocity large enough for shattering. (Note that
graphite grains are not shattered since we only consider
$a\leq 0.25~\mu$m.)
 {Those threshold radii for gyroresonance
are quite robust because they only weakly depend on
the charge, the magnetic field strength, and the
grain density as mentioned in Section \ref{subsubsec:gyro}.}
It is interesting that those grain sizes satisfying
the shattering condition in WNM are nearly the
upper grain size in the Milky Way (MRN). Thus, the upper
limit of the grain size is possibly determined by
shattering in WNM. This issue
is further investigated in Section \ref{subsec:limit}.


Shattering takes place also in CNM for graphite because
graphite has lower shattering threshold velocity than
silicate. However, the result is sensitive to slight
changes in the shattering threshold.
 {Moreover, the grain
velocities acquired by the gyroresonance have uncertainties
coming from the magnetic field strength and the
grain charge, although the uncertainties are generally small
 (Section \ref{subsec:vel}). }
Thus, the shattering in CNM is not
conclusive.
We observe slight coagulation of silicate
with $a\la 10^{-6}~\mathrm{cm}$ in CNM.

In MC and DC, coagulation takes place. In particular,
an appreciable amount of small grains coagulate in DC
because of high density. Since the grain velocities are
lower than the coagulation threshold for
$a\la\mbox{a few}\times 10^{-6}$ cm, the grains accumulates
around $a\sim\mbox{a few}\times 10^{-6}$ cm in DC.
Coagulation occurs up to larger grain radii in DC2 than
in DC1 because the velocity is lower in DC2 than in DC1
because of ion-neutral damping of turbulence
(Section \ref{subsec:vel}).

\subsection{Extinction Curves}\label{subsec:ext}

\begin{figure*}
\begin{center}
\includegraphics[width=7.5cm]{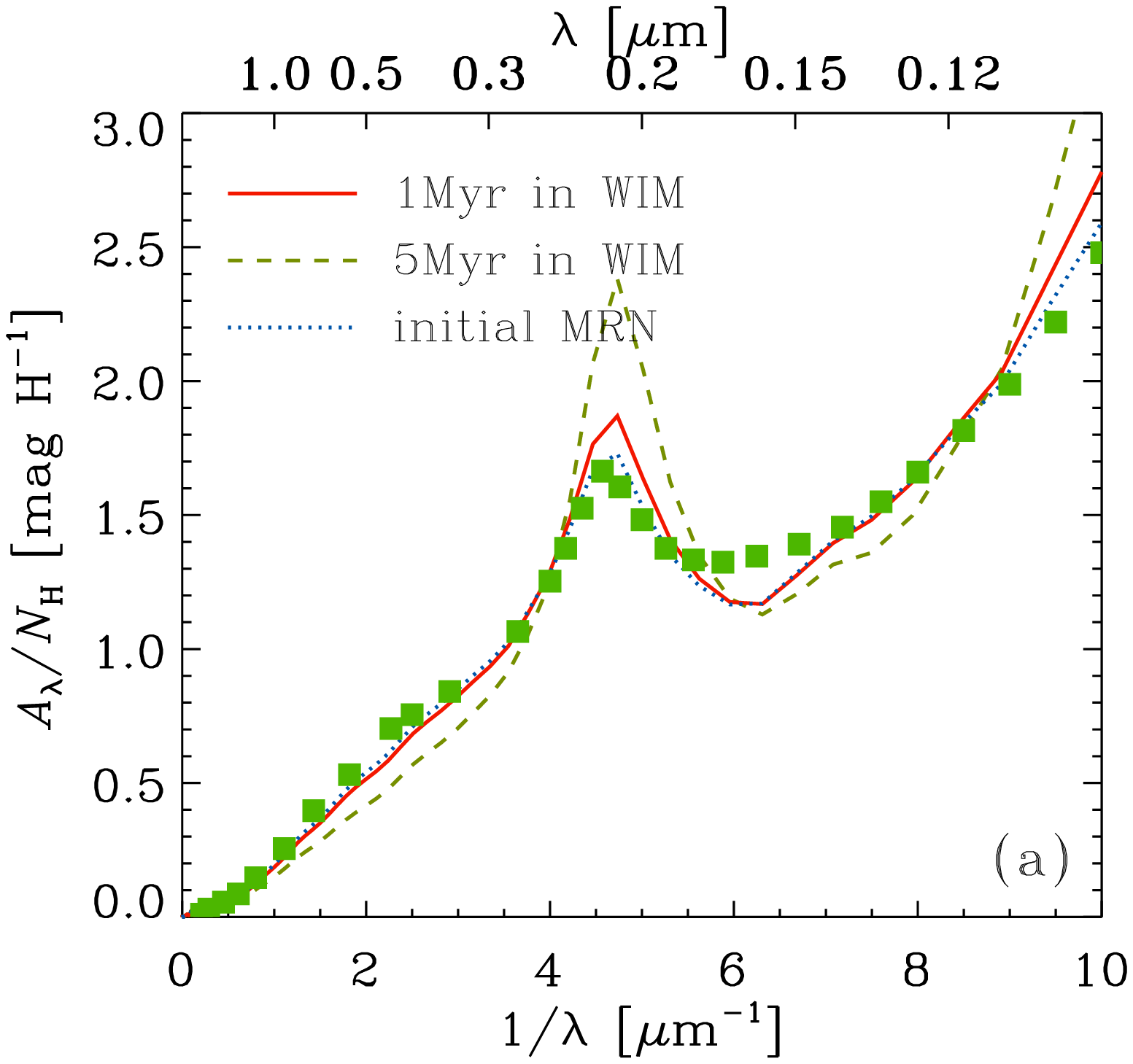}
\includegraphics[width=7.5cm]{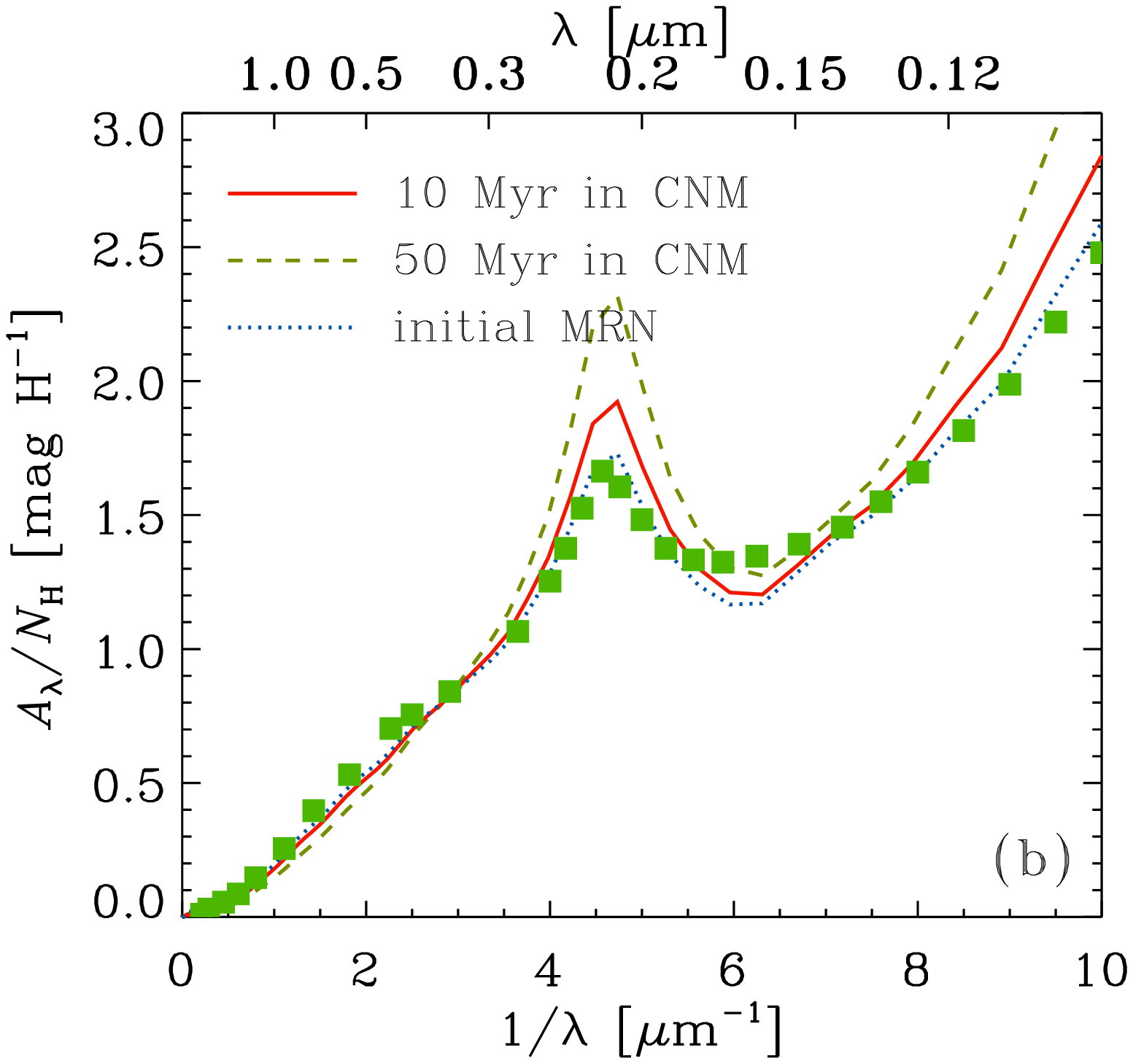}
\includegraphics[width=7.5cm]{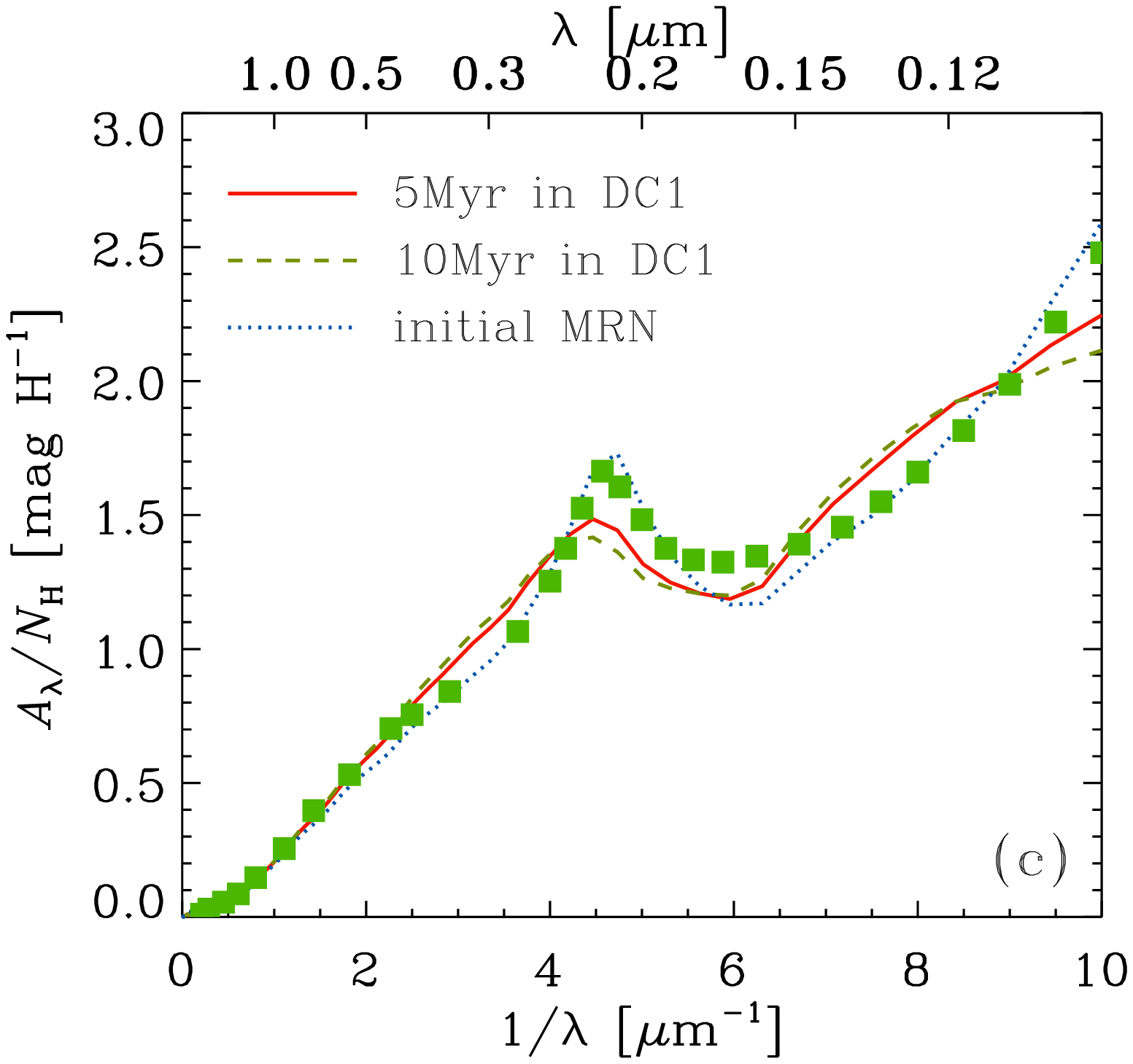}
\includegraphics[width=7.5cm]{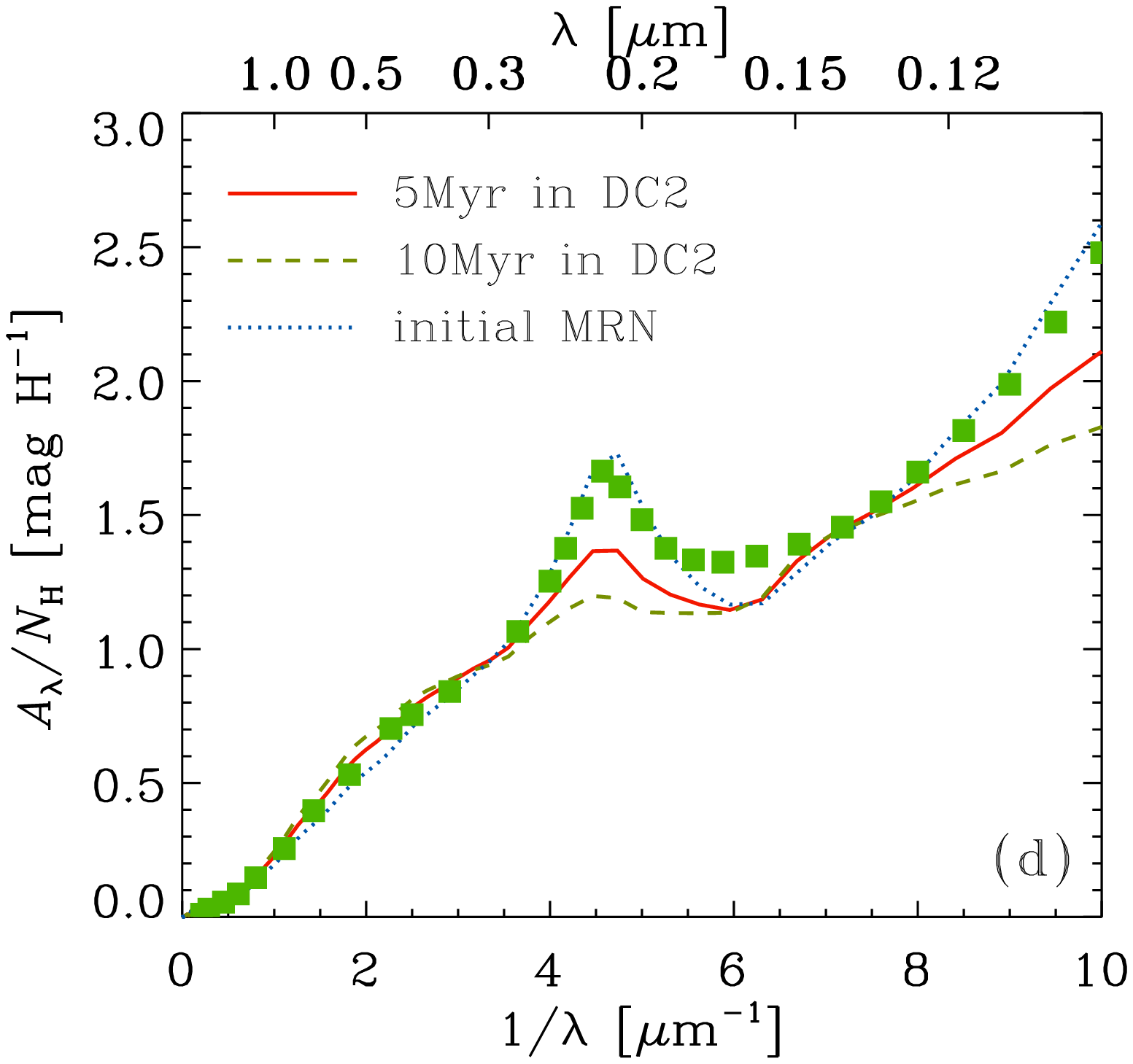}
\end{center}
\caption{The extinction curves of grains processed
in various ISM phases. In each panel, the dotted line
represents the initial MRN distribution.
(a) $t=1$~Myr and 5 Myr in WIM; (b) $t=10$ Myr and 50 Myr
in CNM; (c) $t=5$ Myr and 10 Myr in DC1; and $t=5$ Myr
and 10 Myr in DC2 for the solid and dashed lines,
respectively. The squares show the data of the observed
Milky extinction curve by \citet{pei92}. In this
paper, the extinction is normalized to the hydrogen
column density.
\label{fig:extinc}}
\end{figure*}

For observational comparisons, we calculate the extinction
curves with the method described in
Section \ref{subsec:ext_curve}. In WNM and MC, the grain
size distributions are modified too slightly to change
the extinction curves significantly. The interesting
cases are WIM, CNM, and DC, for which we show the
results in Fig.\ \ref{fig:extinc}.
First of all, the initial MRN distribution reproduces the
Milky Way extinction curve including the UV slope and
the 2175 \AA\ bump. The only deviation is seen
at $1/\lambda\simeq 6~\mu\mathrm{m}^{-1}$. The same deviation
is also seen in \citet{pei92}. Since the aim of this
paper is not precise fitting of the extinction curve,
we do not fine-tune the grain size distribution.
Examples of detailed fitting of the extinction curve can
be seen in \citet{kim94} and \citet{weingartner01}.
Below we
describe some features in the extinction curves
calculated for WIM, CNM, and DC.

\begin{figure*}
\begin{center}
\includegraphics[width=6.5cm]{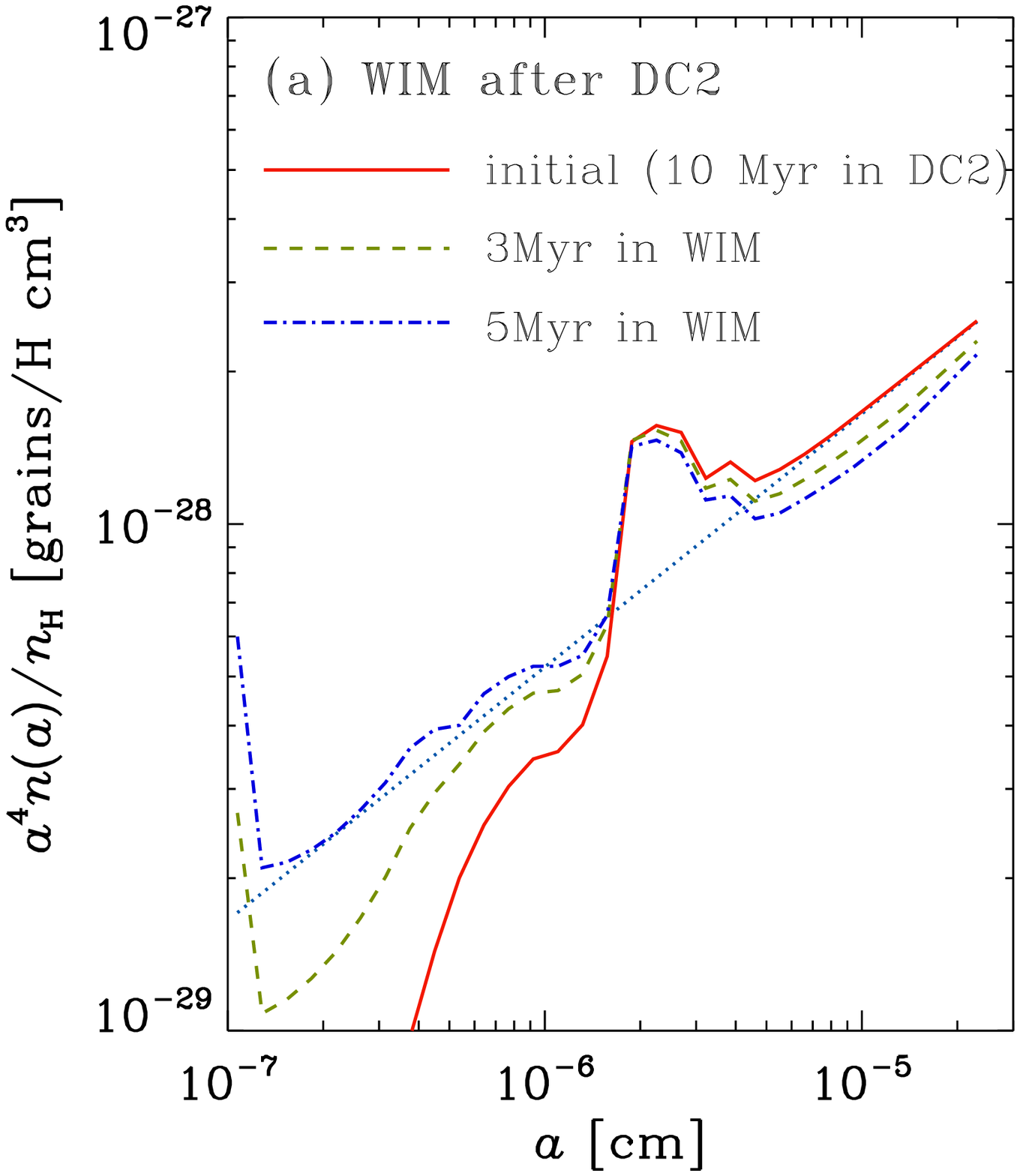}
\includegraphics[width=6.5cm]{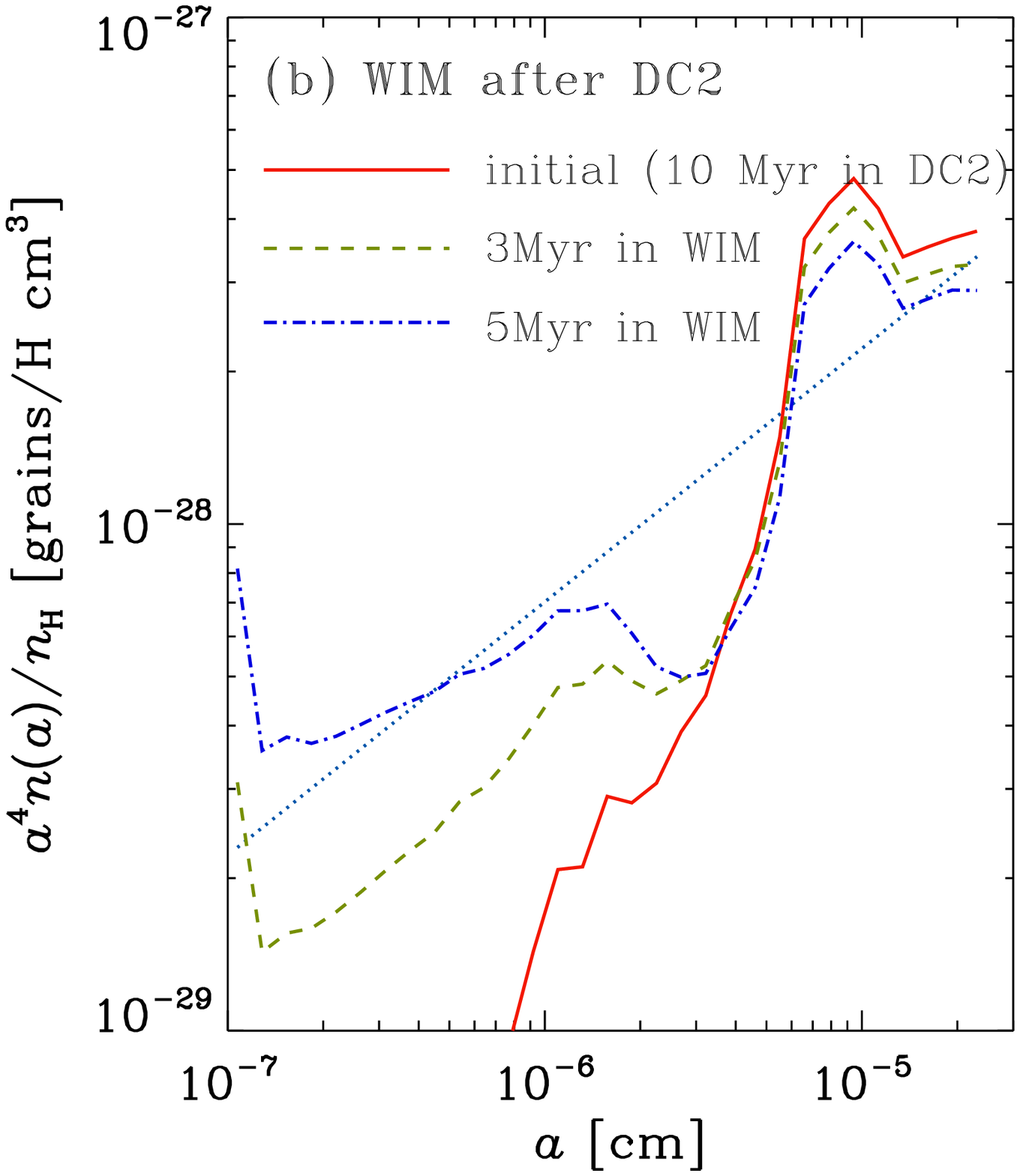}
\end{center}
\caption{The grain size distributions of (a) silicate and (b)
graphite after $t=3$ Myr (dashed line) and $t=5$ Myr (dot-dashed
line) in WIM from the initial distribution (solid line), for
which the size distribution processed for 10 Myr
in DC2 is adopted. The dotted line in each panel shows the
MRN distribution.
\label{fig:dc_wim}}
\end{figure*}

\subsubsection{WIM}\label{subsubsec:wim}

The extinction curves of the grains processed in WIM are
shown in Fig.\ \ref{fig:extinc}a. At $t=5$ Myr, the
2175 \AA\ bump is too high and the UV slope is too steep to
be consistent with the observed Milky Way extinction
curve. Thus, we can conclude that the grains are
continuously processed in WIM for no longer than 5 Myr.
It is interesting to
point out that this timescale is roughly comparable to
the recombination timescale of gas with density
$\sim 0.1~\mathrm{cm}^{-3}$ and temperature $\sim 10^4$ K
($\sim 10^6$ yr; \citealt{spitzer78}) and
to the typical lifetime of massive stars (source of
ionizing photons).

\subsubsection{CNM}\label{subsubsec:cnm}

In Fig.\ \ref{fig:extinc}b, we show the extinction curves
in CNM. Because large graphite grains are shattered,
the 2175 \AA\ bump becomes high and the UV slope becomes
steep. The extinction curves after
shattering in CNM do not deviate largely from the observed
Milky Way extinction curve within 10 Myr. If grains suffer
a longer time of shattering, the extinction curve shows
too high a 2175 \AA\ bump and too steep a UV slope to
be consistent with the observed
extinction curve. However, as noted in
Section \ref{subsec:size_result}, the arguments here are
sensitive to the prediction of grain velocities and the
assumed value of shattering threshold velocities.

\subsubsection{DC}\label{subsubsec:dc}

In Figs.\ \ref{fig:extinc}c and d, we present the
extinction curves in DC1 and DC2, respectively. Because
the grain size is biased toward large sizes after
coagulation, the 2175 \AA\ bump is lower and the UV slope
is less steep than the original curve predicted from the MRN
size distribution.

The extinction curves in DC2 are more consistent with
the observed Milky Way extinction curves than those in
DC1 in the following two points. First, the wavelength
at the peak of 2175 \AA\ bump changes less in DC2 than in
DC1. The observed central wavelengths of 2175 \AA\ bump
in various lines of sight in the Milky Way are
insensitive to the variation of  bump strength
\citep{cardelli89}. The different behaviours of the
2175 \AA\ bump between DC1 and DC2 come from the
``smoothness'' of the size distribution around
$a\sim\mbox{a few}\times 10^{-6}$ cm: In
DC1 the graphite size distribution show a very
steep depletion of grains at $a<3\times 10^{-6}$ cm,
while in DC2 the depletion of such small grains is not
so drastic as in DC1.

Second, the behaviours of $R_V$ in terms of the
2175 \AA\ bump and the UV slope are more consistent with
the observed Milky Way extinction curves in DC2 than in
DC1. The observed Milky Way extinction
curves show that a large $R_V$ is related to
a weak 2175~\AA\ bump and a shallow UV slope
\citep{cardelli89}.
Starting from 3.6 at $t=0$, $R_V$ changes to 3.2
($t=5$ Myr) and 3.1 ($t=10$ Myr) in DC1, while
it changes to 4.2 and 4.8 in DC2. Thus, DC1 has
a trend opposite to the observed one, while DC2
reproduces a right trend between $R_V$, the 2175 \AA\
bump, and the UV slope. Only the grains
with $a<4\times 10^{-6}$ cm, whose velocity is below
the coagulation threshold velocity, can coagulate
in DC1. Since the grain
size change in this small size range does not
affect the extinction in long wavelengths such as
$B$ and $V$ bands,\footnote{From the knowledge of
Mie theory, if the grain size is much smaller
than $\lambda /2\pi$, the extinction
becomes inefficient, i.e.\ $Q\ll 1$, where
$Q$ is the extinction cross section normalized to
the geometrical
cross section \citep{bohren83}.} coagulation of
larger grains
is necessary to change $R_V$. This is why $R_V$ changes
only a little in DC1. In DC2, coagulation
to larger grain sizes indeed occurs and $R_V$
increases as coagulation proceeds.
Considering that there are uncertainties in the threshold
velocity for coagulation and in the grain velocities,
the success in reproducing qualitatively the trend among
$R_V$, the 2175 \AA\ bump, and the UV slope in DC2 supports
the view that coagulation induced by turbulent motions in
dense environments really occurs in the ISM.

\section{Discussion}\label{sec:discussion}

The important features found in the previous section can be
summarized as follows.

(i) The largest effect of shattering is seen in WIM, where
grains with $a\ga 10^{-6}$ cm are efficiently shattered.

(ii) The largest effect of coagulation is observed in DC
around $a\la\mbox{a few}\times 10^{-6}$ cm.

(iii) Grains with $a\ga\mbox{a few}\times 10^{-5}$ cm can be
shattered in WNM and graphite grains with $a\ga 10^{-5}$ cm
may be quite efficiently destroyed in CNM. These destructions
could affect the upper limit of grain size in ISM.

(iv) On the other hand, the lower limit of grain size may be
determined by coagulation in DC and MC.

The features (i) and (ii) indicate that once grains are
included in WIM or DC, the grain size distribution is
significantly modified. It is interesting to note that
the shattered grains in (i) and the coagulated grains in
(ii) have a similar size. Thus, it is worth investigating if
the MRN size distribution can be realized as a balance
between (i) and (ii). This point is investigated in
Section \ref{subsec:equilibrium}.

Regarding the feature (iii), as mentioned in
Section \ref{subsubsec:cnm}, the results in CNM are
sensitive to the grain velocities and the shattering
thresholds. We leave more careful treatment of shattering
in CNM for future work.
Shattering in WNM is interesting to investigate,
since turbulence in WNM accelerates grains with
$a\ga \mbox{a few}\times 10^{-5}$ cm much above the
threshold velocity for shattering. This size really
matches the upper limit of the grain size distribution
(MRN). This point is investigated in
Section \ref{subsec:limit}.

The issue (iv) has already been investigated and discussed
in Section \ref{subsubsec:dc}.

\subsection{Grain size distributions in
diffuse-dense phase exchange}\label{subsec:equilibrium}

In ISM, mass is exchanged between various phases
\citep{mckee77,ikeuchi88}. Thus, it is important to
investigate the effects of multi-phase ISM on the evolution
of grain size distribution, although the main aim of this
paper is to examine the dust processing in individual phases.
The largest shattering and coagulation effects are seen in WIM
and DC, respectively, and we here examine the dust processing
in both WIM and DC to address a possible importance of
multi-phase ISM in determining the grain size distribution.
For DC, we adopt DC2 because of the success in explaining the
trend of $R_V$ in terms of the UV slope and the
2175~\AA\ bump (Section \ref{subsubsec:dc}).

We start from the size distribution of grains processed in
DC2 for 10 Myr. Then, we apply the condition of WIM.
In Fig.\ \ref{fig:dc_wim}, we show the results at
$t=3~\mathrm{Myr}$ and 5 Myr in WIM. Around 5 Myr, the number
of small grains is recovered to the level of the MRN
distribution. In other words, if grains pass their lifetimes
in WIM more than in DC, the grains are shattered too much
to be consistent with the MRN distribution. This implies a
short lifetime of WIM. Combining this short lifetimes
of WIM with a theoretically implied timescale for the phase
exchange ($\mbox{a few}\times 10^7$--$10^8$ yr;
Section \ref{subsec:timescale}), we obtain a picture that a
large fraction of warm medium is
in a neutral form and a certain small fraction is ionized.
It is interesting to point out that such a short timescale
is consistent with the recombination timescale as
mentioned in Section \ref{subsubsec:wim}.

\begin{figure}
\begin{center}
\includegraphics[width=7.5cm]{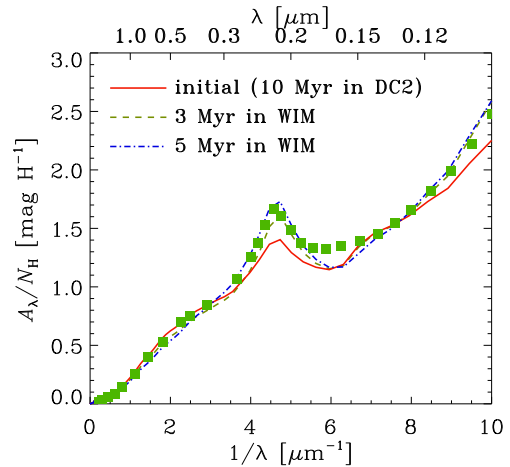}
\end{center}
\caption{The extinction curves calculated for the size
distributions in Fig.\ \ref{fig:dc_wim}. The solid line
represents the initial size distribution (10 Myr
in DC2), and the dashed and dot-dashed lines show the
extinction curves at $t=3$ Myr and 5 Myr in WIM,
respectively.
\label{fig:extinc_exchange}}
\end{figure}

The corresponding extinction curves are shown in
Fig.\ \ref{fig:extinc_exchange}. The Milky Way extinction
curve is indeed recovered by the phase exchange. This
demonstrates that it is really possible to reproduce the
Milky Way extinction curve by considering dust grains
processed in multiphase medium.

The above phase exchange model is too simple, and the
realistic ISM has more continuous density distribution
and more complicated structure of turbulence \citep{wada01}.
Such complexity should tend to eliminate the specific
features such as accumulation of grains around
$a\sim \mbox{a few}\times 10^{-6}$ cm in DC and
selective grain destruction at $a\sim 10^{-6}$ cm in
WIM. Thus, we expect that the grain size distribution
becomes smoother in realistic ISM than we calculate in
this paper.

\begin{figure*}
\begin{center}
\includegraphics[width=6.5cm]{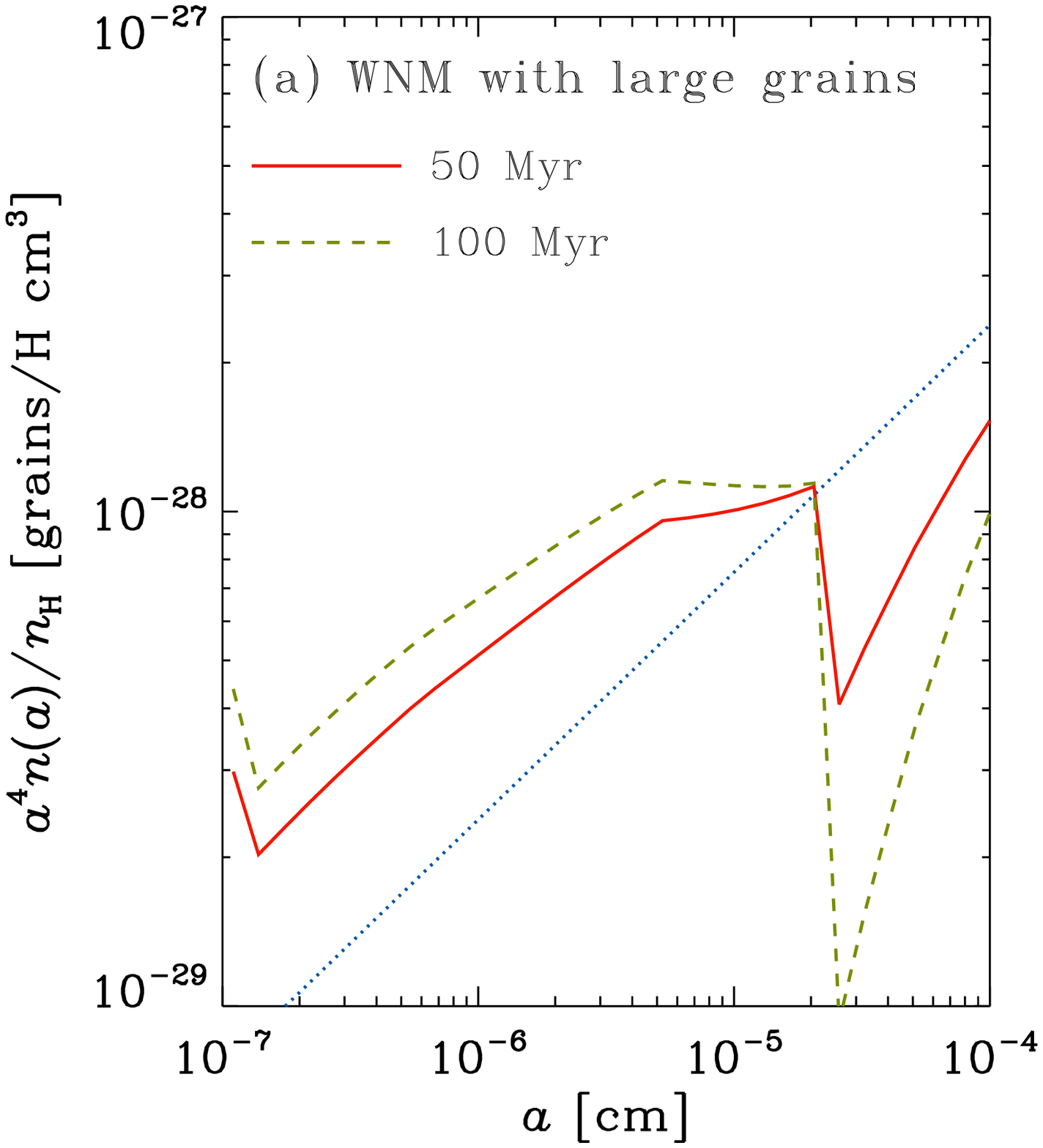}
\includegraphics[width=6.5cm]{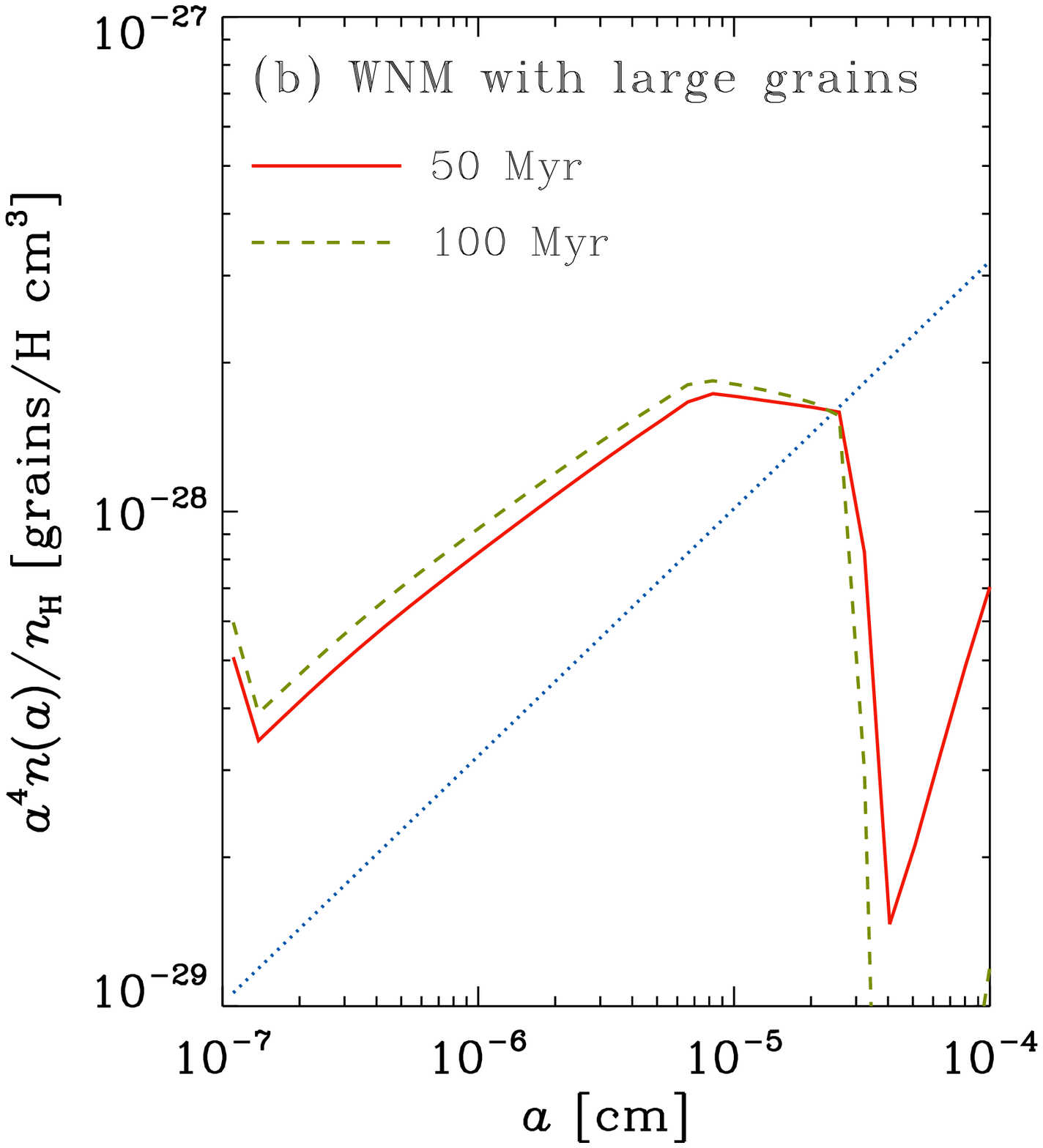}
\end{center}
\caption{The grain size distributions of (a) silicate
and (b) graphite after $t=50$ Myr (solid line) and
$t=100$ Myr (dashed line) in WIM
for the MRN size distribution
extending up to $a=10^{-4}$ cm as the initial condition
(dotted line).
\label{fig:wnm_large}}
\end{figure*}

\subsection{Upper and lower limits of grain size}
\label{subsec:limit}

According to MRN, the upper grain radius is
$\sim 0.25~\mu$m (see also \citealt{kim94}). Coagulation
has negligible influence on grains larger than
$a\sim 0.2~\mu$m both for silicate and for graphite because
they generally obtain larger velocity than the coagulation
thresholds. Thus, if there is no grain with
$a\ga 0.2~\mu$m initially, it is not possible to
make such large grains by coagulation in ISM.

Even if grains larger than $a\sim 0.2~\mu$m form by
condensation in stellar ejecta, shattering could destroy
such large grains. \citet{nozawa03} show that silicon grains
with $a>0.2~\mu$m form in Type II supernovae. In the
outflows from evolved late-type stars, the grain radius is
expected to become of order $\sim 0.1~\mu$m \citep{gail99},
and grains with $a\ga 0.2~\mu$m may have a chance to form.
It is interesting to note that grains with
$a\ga 0.2$--0.3 $\mu$m are accelerated above the
shattering threshold in WNM. Thus, shattering in
WNM may play a central role in determining the upper limit
of the grain size in ISM.

In order to examine whether or not shattering in WNM
really plays a role in determining the upper limit of the
grain size, we perform a test by adopting an initial grain
size distribution extending up to $a=1~\mu$m with the total
mass of grains conserved. Then the evolution of the grain
size distribution is calculated by applying the conditions
in WNM. Fig.\ \ref{fig:wnm_large} shows the results. We
observe that the grains with $a\ga 0.2$--0.3 $\mu$m
are significantly shattered in 50 Myr. Thus, shattering in
WNM is a strong candidate for the determining mechanism of
the upper limit of grain size.

\begin{figure}
\begin{center}
\includegraphics[width=7.5cm]{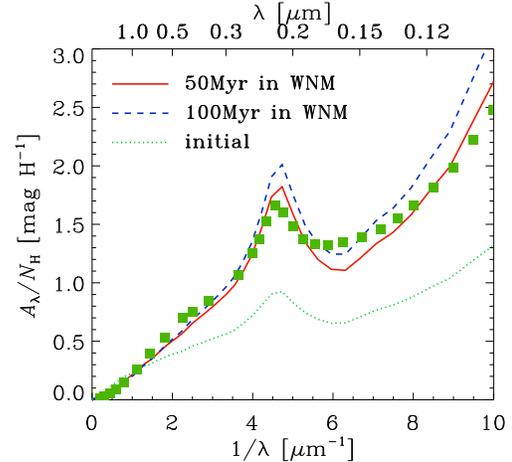}
\end{center}
\caption{The extinction curves calculated for the size
distributions in Fig.\ \ref{fig:wnm_large}. The dotted
line represents the initial extinction curve (MRN size
distribution extending up to $a=10^{-4}$ cm). The solid
and dashed lines show the extinction curves of grains at
$t=50$ Myr and $t=100$ Myr in WNM, respectively.
\label{fig:extinc_large}}
\end{figure}

In Fig.\ \ref{fig:extinc_large}, we show the corresponding
extinction curves. The initial extinction curve is
significantly lower than the observed one because large
grains tend to have low mass absorption coefficients.
However, after 50 Myr, the level of the extinction is
already consistent with the Milky Way curve. This means
that shattering of large grains in WNM is efficient enough
to reproduce the upper grain size consistent with the
observed Milky Way extinction curve.

\begin{figure*}
\begin{center}
\includegraphics[width=6.5cm]{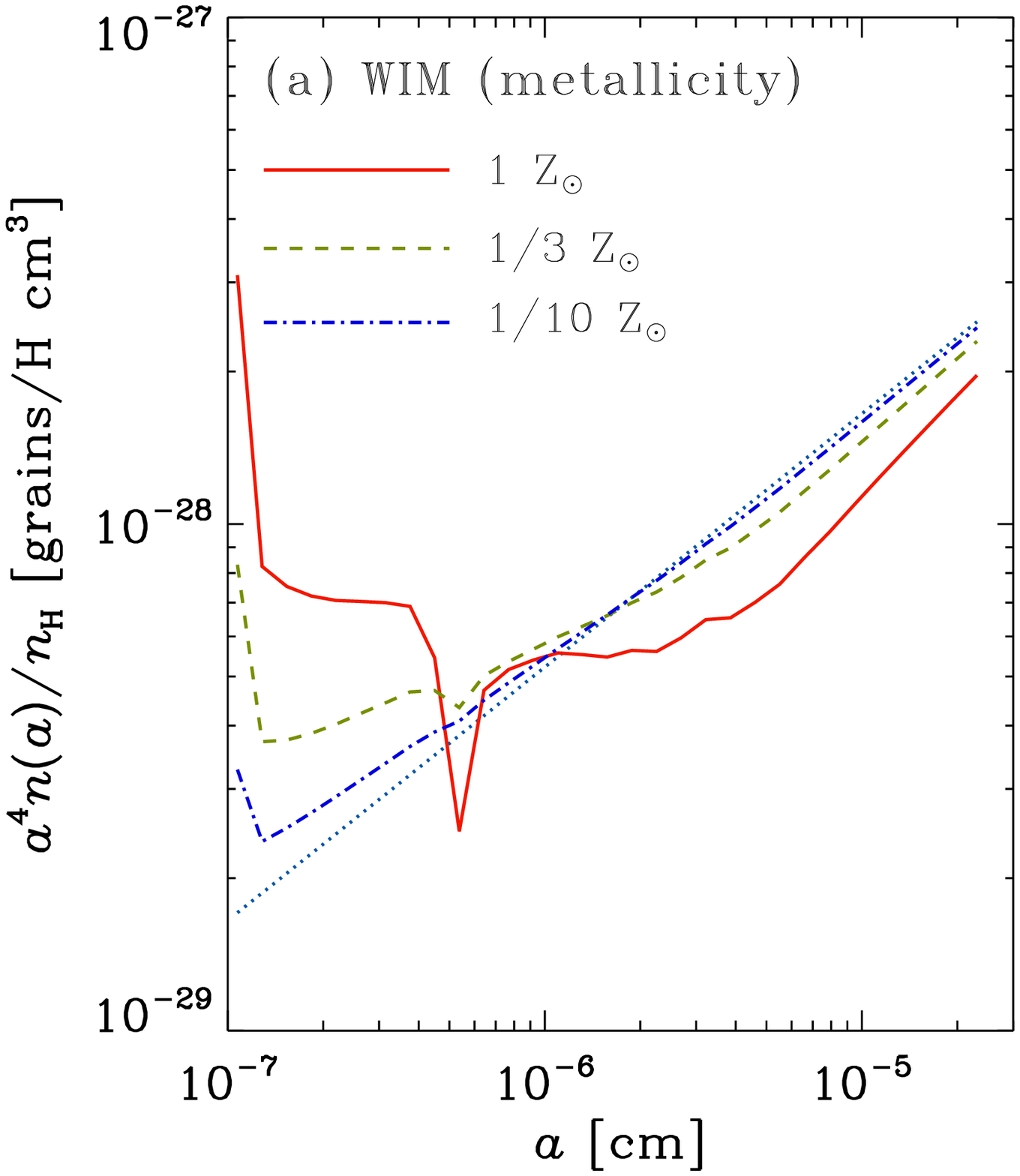}
\includegraphics[width=6.5cm]{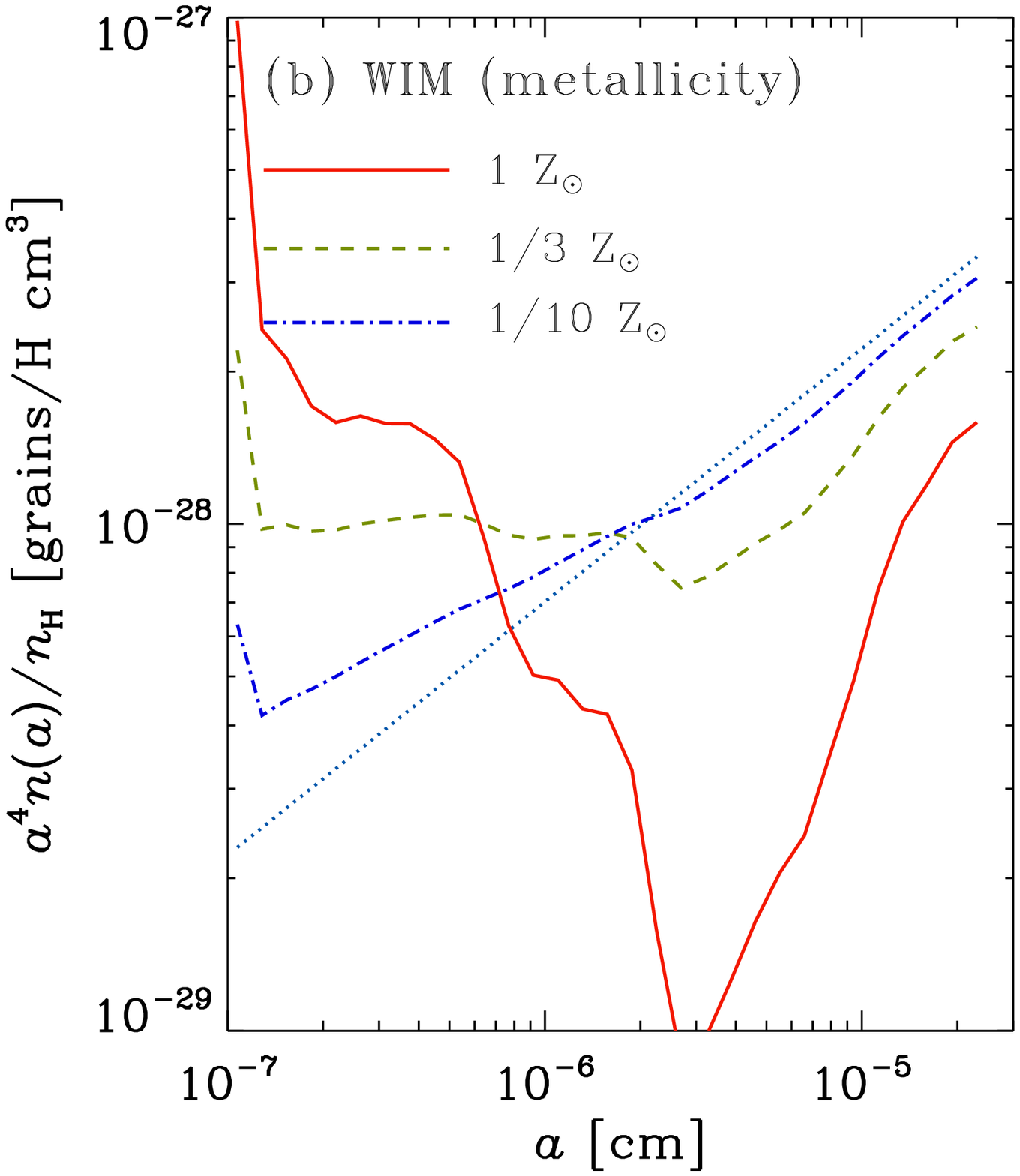}
\end{center}
\caption{The grain size distributions of (a) silicate
and (b) graphite at $t=10$ Myr in WIM. The solid, dashed,
dot-dashed lines show $Z=1$, 1/3 and 1/10~$\mathrm{Z}_\odot$,
respectively. The dust abundance (vertical axis) of the
dashed and dot-dashed lines are multiplied by 3 and 10,
respectively, to offset the low dust abundances.
\label{fig:wim_metallicity}}
\end{figure*}

\begin{figure*}
\begin{center}
\includegraphics[width=6.5cm]{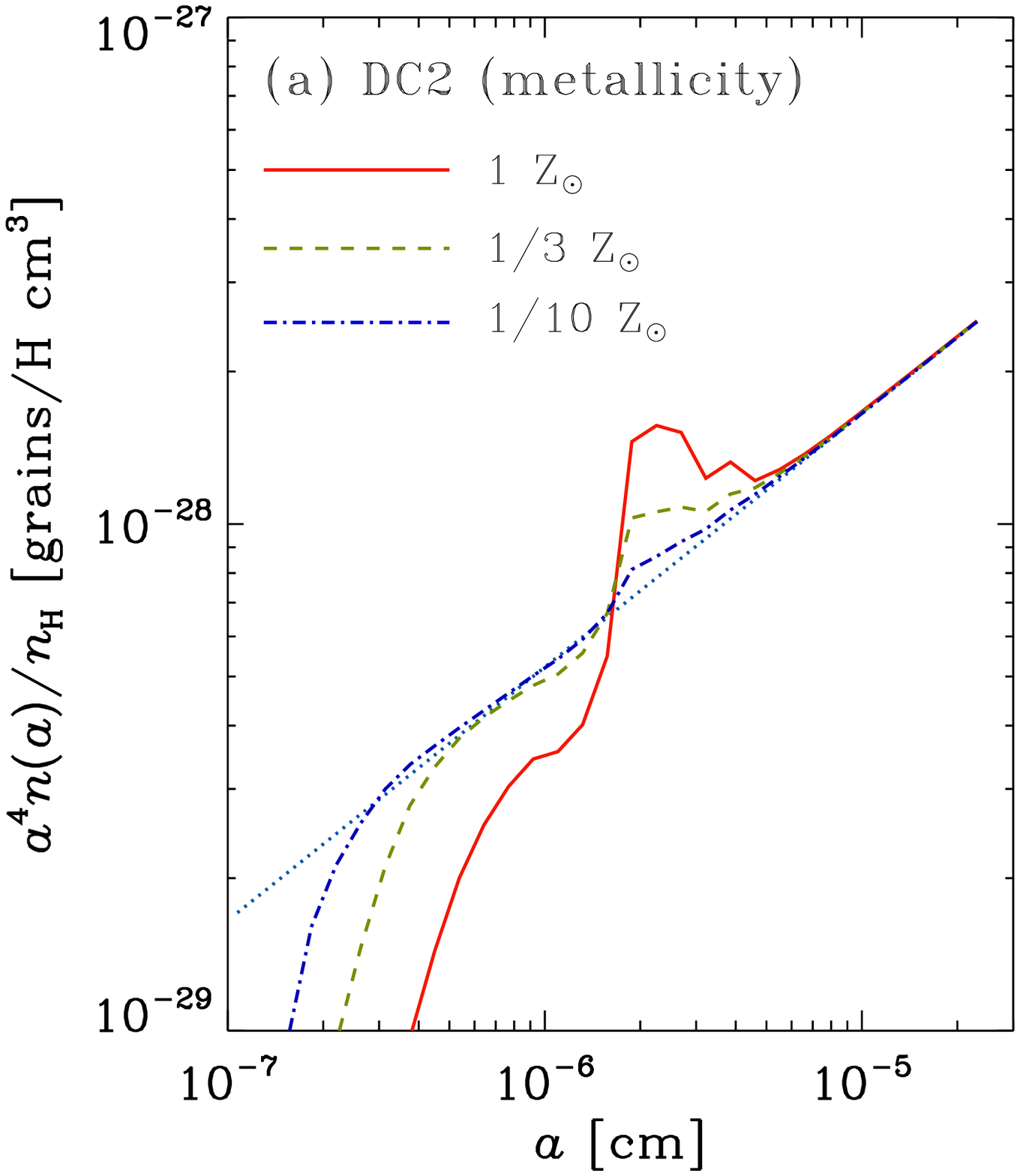}
\includegraphics[width=6.5cm]{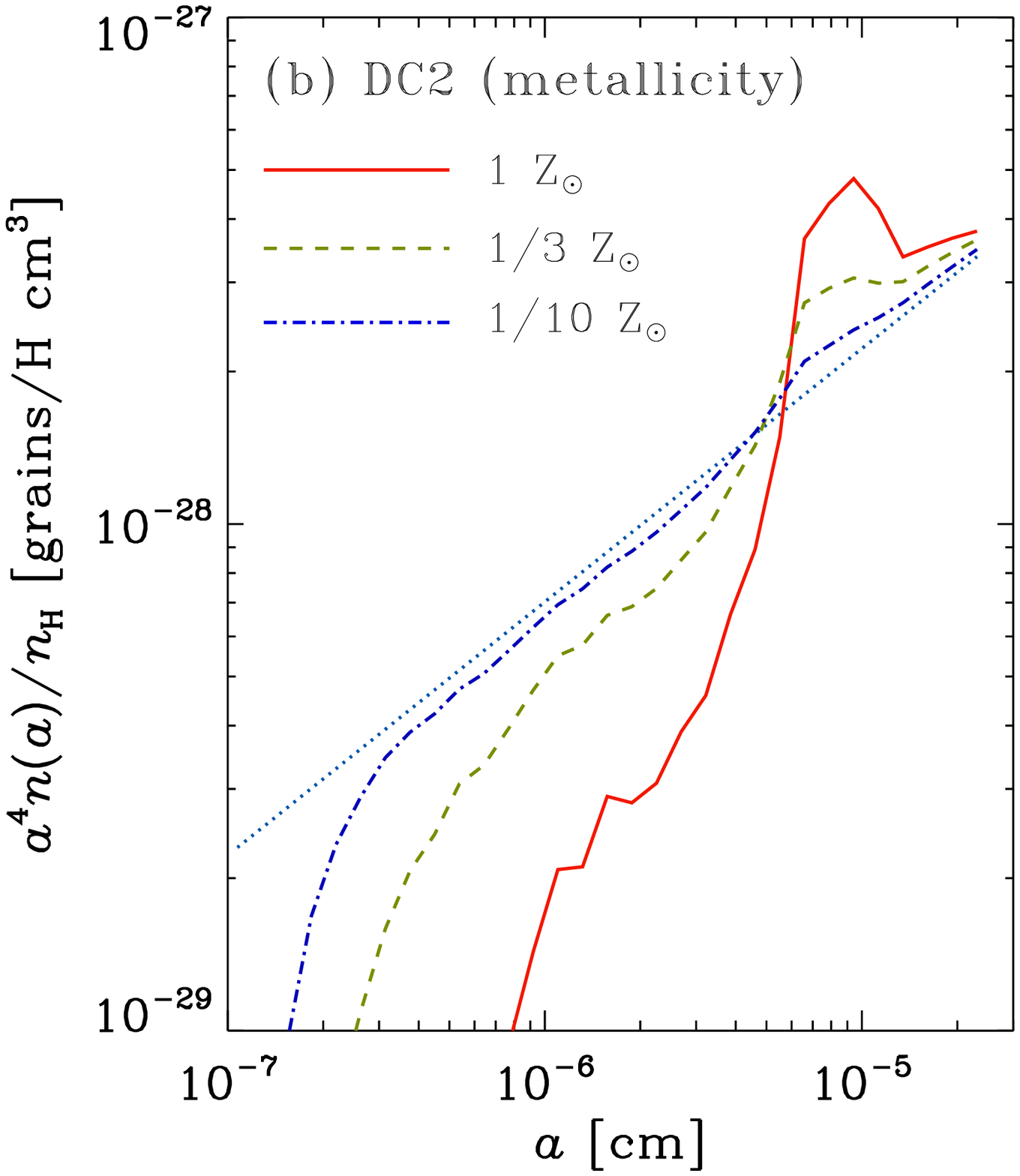}
\end{center}
\caption{Same as Fig.\ \ref{fig:wim_metallicity} but in DC2.
\label{fig:dc_metallicity}}
\end{figure*}

\subsection{In the context of galaxy evolution}

The efficiencies of shattering and coagulation are affected by
the grain abundance. This indicates that
metal-poor galaxies, which are generally poor in dust content
\citep{issa90}, have different grain size distributions.
Here we examine the metallicity dependence of shattering
and coagulation. We assume that the dust-to-gas ratio is
proportional to the metallicity $Z$; that is, we adopt
${\cal R}=4.0\times 10^{-3}Z/\mathrm{Z}_\odot$ and
$3.4\times 10^{-3}Z/\mathrm{Z}_\odot$ for silicate and graphite,
respectively, in equation (\ref{eq:norm_grain}).
In other words, the dust density in the ISM is proportional
to the metallicity, and we expect that the effects of
shattering and coagulation become weak as the metallicity
decreases.
 {The turbulence model and the grain velocities are not
changed, which means that
we implicitly assume that the parameters listed in
Table \ref{tab:yan} are fixed.}

We test WIM and DC, where shattering and coagulation,
respectively, are the most efficient among the various
phases. In Fig.~\ref{fig:wim_metallicity}, we show the
grain size distributions in WIM at $t=10$ Myr. We apply
a longer timescale than adopted in the other part of
this paper to
enhance the effect of shattering. We observe that the
shattering effect is significantly reduced at
$1/10~\mathrm{Z}_\odot$. The same is true for coagulation
in DC as shown in
Fig.\ \ref{fig:dc_metallicity}, where we adopt DC2 because
of the success in reproducing the trend of $R_V$ in terms
of the UV slope and the 2175 \AA\ bump
(Section \ref{subsubsec:dc}).
Thus, as the metallicity decreases, the relative importance
of processing by interstellar turbulence becomes
minor in determining the grain size distribution.
This indicates that the initial grain size distribution at
the grain formation in stellar ejecta is relatively
preserved in metal-poor galaxies
(typically $Z<1/10~\mathrm{Z}_\odot$),
although we should keep in mind that there are
other processes, such as interstellar shocks by supernovae,
which could modify the grain size distribution in any
metallicity.

\subsection{Toward the grain evolution in protoplanetary
discs}

 {The condition of turbulence in the circumstellar discs
is still unclear. Let us consider protoplanetary discs.
According to \citet{nomura06}, turbulence is very weak
($\delta V\sim 0.01\mbox{--}0.1c_\mathrm{s}$, where
$c_\mathrm{s}$ is the sound speed). The acceleration by
turbulence will be marginal in this case, and grain
motions are
more likely to be Brownian. As a result coagulation is at
least as efficient as in DC.  As shown in
Figs.\ \ref{fig:sil} and \ref{fig:gra}, small grains with
$a\la 10^{-6}$ cm are strongly depleted in DC because of
coagulation. 
Thus, we can justify that the grain size distribution in
protoplanetary discs is biased to radii $\ga 10^{-6}$ cm.
Moreover, because grain velocities are
expected to be lower than the coagulation threshold even
at $a>10^{-6}$ cm, grains grow further.}

 {As shown by \citet{sano00}, the grain size in
protoplanetary discs is important in determining the
unstable regions for magnetorotational instability, which
induces MHD turbulence \citep{balbus98}. Consequently the
grain size distribution
is further affected by the presence/absence of
the turbulent motion determined by the instability/stability
condition. The coupling between turbulence and
grain size is interesting to investigate as a future work.}

\section{Summary}\label{sec:sum}

We have investigated the effects of shattering and
coagulation on the dust size distribution in turbulent ISM,
adopting the typical velocities of dust grains as a function
of grain size from YLD04. By using a scheme of grain
shattering and coagulation which we have developed in this
paper based on JTH96 and \citet{chokshi93}, we have calculated
the evolution
of grain size distribution in turbulent ISM. Since large
grains tend to have large velocities because of decoupling
from small-scale turbulent
motions, large grains tend to be shattered. On the other hand,
because of small surface-to-volume ratio, large grains
require more time to be destroyed.

Large shattering effects are indeed seen in WIM for
grains with $a\ga\mbox{a few}\times 10^{-6}$ cm. In the
supernova shocks, such small grains are decelerated
quickly by gas drag and larger grains tend to be shattered
more efficiently (JTH96). Graphite grains are predicted to
be shattered also
in CNM, but the result in CNM is sensitive to the
threshold velocity for shattering. Coagulation
significantly modifies the grain size distribution in DC.
In fact, the correlation among $R_V$, the carbon bump,
and the UV slope in the
observed Milky Way extinction curves is qualitatively
reproduced by the coagulation in DC. We have also shown
that the upper limit of the grain size in ISM can be
determined by the shattering in WNM.

If a large fraction of
ISM experiences either WIM or DC, the grain size
distribution in ISM may be determined by a balance
between shattering in WIM and coagulation in
DC. Considering that the effects of shattering and
coagulation become small in metal-poor environments,
the regulation mechanism of grain size distribution is
quantitatively different between metal-poor and metal-rich
environments.

\section*{Acknowledgments}
We thank the anonymous referee for useful comments which improved
this paper considerably.
We are grateful to A Lazarian for reading the manuscript and his
suggestions and M. Umemura and R. Nishi for helpful discussions.

\end{document}